%% file: main.tex
\documentclass[manuscript,screen,balance=false,nonacm]{acmart}

\renewcommand{\vec}[1]{\mathbf{#1}}
\AtBeginDocument{%
  }

\setcopyright{acmlicensed}
\copyrightyear{2018}
\acmYear{2018}
\acmDOI{XXXXXXX.XXXXXXX}

\acmConference[Conference acronym 'XX]{Make sure to enter the correct
  conference title from your rights confirmation emai}{June 03--05,
  2018}{Woodstock, NY}
\acmISBN{978-1-4503-XXXX-X/18/06}

\usepackage{graphicx}
\usepackage{multicol}
\usepackage{multirow}
\usepackage{makecell}
\usepackage{wrapfig}
\usepackage{enumitem}
\usepackage{threeparttable}
\usepackage{amsmath}
\usepackage{physics}
\begin{document}
\title{A Survey on Time-Series Distance Measures}


\author{John Paparrizos}
\affiliation{%
  \institution{The Ohio State University}
  \city{Columbus}
  \country{USA}}
\email{paparrizos.1@osu.edu}

\author{Haojun Li}
\affiliation{%
  \institution{The Ohio State University}
  \city{Columbus}
  \country{USA}}
\email{li.14118@.osu.edu}

\author{Fan Yang}
\affiliation{%
  \institution{The Ohio State University}
  \city{Columbus}
  \country{USA}}
\email{yang.7007@osu.edu}

\author{Kaize Wu}
\affiliation{%
  \institution{University of Chicago}
  \city{Chicago}
  \country{USA}}
\email{kaizewu@uchicago.edu}

\author{Jens E. d'Hondt}
\affiliation{%
    \institution{Eindhoven University of Technology}
    \city{Eindhoven}
    \country{the Netherlands}}
\email{j.e.d.hondt@tue.nl}

\author{Odysseas Papapetrou}
\affiliation{%
    \institution{Eindhoven University of Technology}
    \city{Eindhoven}
    \country{the Netherlands}}
\email{o.papapetrou@tue.nl}
\renewcommand{\shortauthors}{Paparrizos et al.}

\begin{abstract}
  Distance measures have been recognized as one of the fundamental building blocks in time-series analysis tasks, e.g., querying, indexing, classification, clustering, anomaly detection, and similarity search. The vast proliferation of time-series data across a wide range of fields has increased the relevance of evaluating the effectiveness and efficiency of these distance measures. To provide a comprehensive view of this field, this work considers over 100 state-of-the-art distance measures, classified into 7 categories: lock-step measures, sliding measures, elastic measures, kernel measures, feature-based measures, model-based measures, and embedding measures. Beyond providing comprehensive mathematical frameworks, this work also delves into the distinctions and applications across these categories for both univariate and multivariate cases. By providing comprehensive collections and insights, this study paves the way for the future development of innovative time-series distance measures.
\end{abstract}




\received{20 February 2007}
\received[revised]{12 March 2009}
\received[accepted]{5 June 2009}

\maketitle

\input{section/S01_introduction}
\input{section/S02_preliminary}
\input{section/S03_taxonomy}
\input{section/S04_lockstep}
\input{section/S05_elastic}
\input{section/S06_sliding}
\input{section/S07_kernel}
\input{section/S08_feature}
\input{section/S09_model}
\input{section/S10_embedding}
\input{section/S11_mtsdist}
\input{section/S13_conclusion}

\bibliographystyle{ACM-Reference-Format}
\bibliography{Survey_Bib}


\end{document}

%% file: section/S01_introduction.tex
\section{Introduction}

With the advancement of techniques in sensing, networking, storage, and data processing, it has become feasible to collect, store, and process massive collections of measurements over time \cite{raza2015practical,keogh2006decade,paparrizos2021vergedb, paparrizos2018fast, liu2024adaedge}, referred to as \textit{time series}. Time series analysis, with its ability to capture temporal relationships between data points, has attracted significant interest across various academic and industrial domains \cite{mckeown2016predicting, paparrizos2016screening,paparrizos2016detecting,jeung2010effective,goel2016social,zupko2023modeling, YANG2024101641, li2023towards-noise} such as electrical engineering \cite{kashino1999time,raza2015practical}, astronomy \cite{alam2015eleventh,wachman2009kernels}, finance \cite{brockwell2016introduction,gavrilov2000mining}, energy \cite{bach2017flexible,alvarez2010energy}, environment \cite{goddard2003geospatial,webster2005changes,hoegh2007coral,morales2010pattern}, bioinformatics \cite{ernst2006stem,bar2004analyzing,bar2012studying}, medicine \cite{richman2000physiological,costa2002multiscale,peng1995quantification}, and psychology \cite{kasetty}. 
In various domains and Internet-of-Things applications, the increasing volume of time series data has prompted the need for efficient techniques in data processing and analysis \cite{jiang2020pids, jiang2021good,liu2021decomposed, liu2023amir, krishnan2019artificial, dziedzic2019band}. 

Distance functions, as one of the fundamental building blocks in time-series analysis, are engineered to define dissimilarity between signals \cite{paparrizos2023querying,d2024beyond}.
They have been widely used in every time-series downstream task such as similarity search \cite{agrawal1993efficient, ding2008querying}, indexing \cite{cai2004indexing, echihabi2018lernaean, paparrizos2022fast}, clustering \cite{alon03, kalpakis2001distance,paparrizos2015k,paparrizos2017fast,bariya2021k,paparrizos2023odyssey}, classification \cite{lines2015time,ratanamahatana2004making, paparrizos2019grail} and anomaly detection \cite{breunig2000lof, dallachiesa2014top, boniol2021sand, boniol2021sandinaction, sylligardos2023choose, paparrizos2022tsb, paparrizos2022volume, boniol2022theseus, boniol2023new, liu2024time, liu2024elephant,boniol2024adecimo,boniol2024interactive}. 
Understanding the relationships between instances, such as similarities or dissimilarities, offers valuable insights for uncovering intra-class and inter-class patterns across different domains. Since the notion of similarity is highly dependent on the context of the data and downstream tasks, a large diversity in distance measures has emerged. Different measures have been created to capture diverse notions of similarity, where the degree of similarity between two time series can vary greatly between any two measures. 

However, determining the dissimilarity of two time series is not a trivial task. In real-world applications, time series data can be distorted in various ways, which increases the challenge in the analysis process \cite{paparrizos2015k}. Distortions can include (1) scaling and translation, where two time series have different amplitudes (scaling) and/or offsets (translation); (2) shifting, where two sequences have different phases and alignment should be considered; (3) occlusion, where some subsequences of time series are missing in the dataset; (4) uniform scaling, where time series have different lengths; (5) complexity, where time series with similar shape exhibit complexity at different levels, e.g., one time series could suffer from more noise perturbation while the other may experience less. Neglecting these distortions can result in practical issues, underscoring the importance of addressing them properly. To illustrate this, consider the classic measure of Euclidean Distance (ED), which evaluates the similarity of each element of a time series compared to the corresponding element at the same time point in another series. We will refer to measures of this type as \textit{lock-step}. Such measure design may suffer significant performance degradation when two compared time series exhibit the same pattern at a different temporal point (shift-distortion). 
Additionally, various distortions across time steps and dimensions can create further complications in the process. In special cases, significant noise levels can make it difficult to extract meaningful information from the raw data. 
These issues highlight the need for proper data normalization, which functions as one crucial preprocessing component to alleviate the noise issue, e.g., scaling and translation distortions. 

\begin{figure}
\centering
\includegraphics[width=\linewidth]{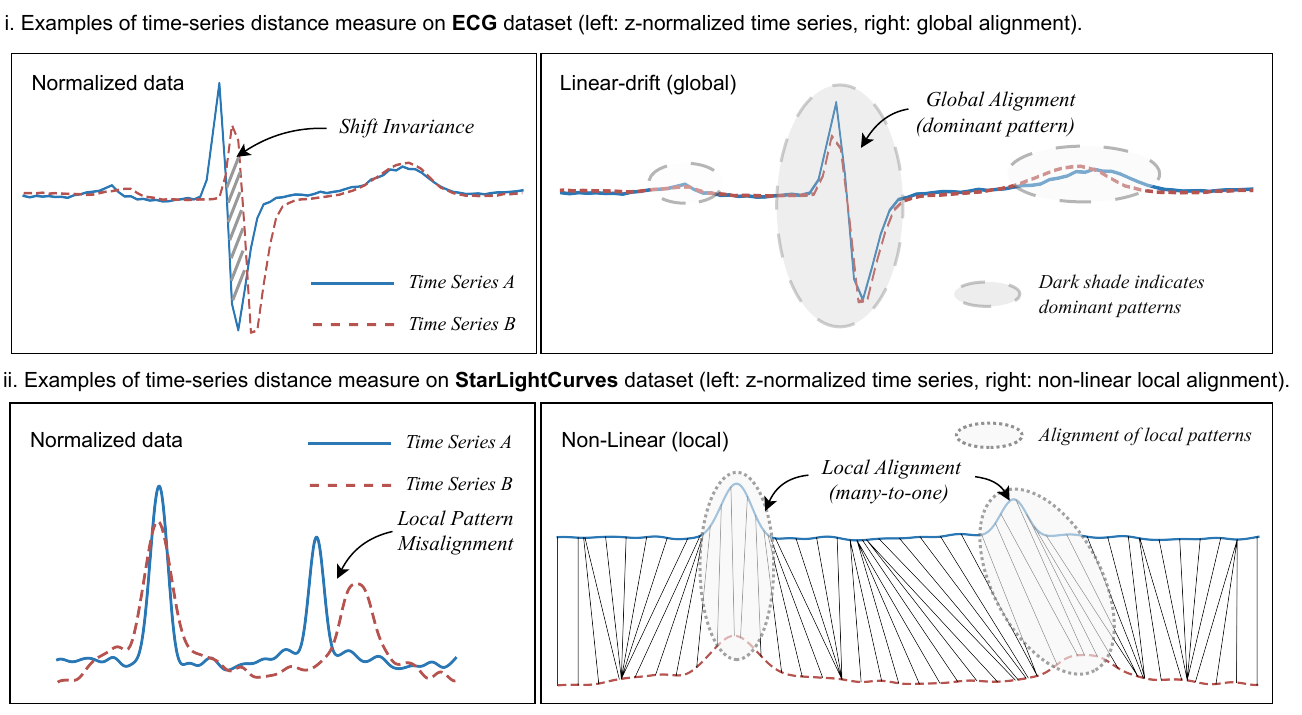} \\
\caption{Overview of the time-series distance measure. Left: normalized time series examples. Right: two different alignment strategies.
}
\label{fig:intro}
\vspace{-0.3cm}
\end{figure}

For those distortions that cannot be addressed in the preprocessing stage, 
strategies like elaborate time series alignment have shown significant benefits in the distance measure design. 
For example, Dynamic Time Warping (DTW) \cite{berndt1994using} utilizes dynamic programming to find the optimal alignment between two time series. Compared with conventional lock-step measures,  these ``elastic'' characteristics enable the distance measure to capture not only one-to-one mapping but also one-to-many mapping across time steps.
However, this mapping process may lead to high complexity, i.e., $\mathcal{O}(n^2)$ time with time series length $n$, which hinders its application in time-sensitive tasks. 
To reduce the time complexity,
Shape-based Distance (SBD) \cite{paparrizos2015k}, as one of the sliding measures, efficiently aligns two time series by leveraging the merit of Fast Fourier Transform (FFT). This approach reduces the time complexity to $\mathcal{O}(n\log(n))$.  Figure \ref{fig:intro} visualizes the distinctions between these two aforementioned alignment strategies. Numerous other strategies have been proposed to address the issues in different ways, which will be described in the following sections.

As researchers and practitioners in this space, these considerations imply a search process through many available measures to discover the measure giving the best results for a particular problem. However, the process of choosing a correct measure can be laborious and time-consuming. The above concerns are then compounded by many measures, including parameters that require careful selection or tuning. In many cases, an exhaustive search is not desirable or is practically infeasible. Therefore, practitioners must be guided by broad knowledge of time series distance measures and their relative strengths and weaknesses.




Following the proliferation of sources and applications involving time series, in the past couple of decades, many studies and surveys attempted to capture and categorize the state-of-the-art methods \cite{wang2013experimental,liao2005clustering,aghabozorgi2015time,cha,dongpu2021survey,shifaz23,john2020debunking,ding2008querying,giusti2013empirical}. ~\cite{cha}  categorizes over 50 lock-step measures between probability density
functions. Ding et al.~\cite{ding2008querying} evaluate 8 representation methods and 9 similarity measures across various application domains, demonstrating the effectiveness of each. Giusti and Batista~\cite{giusti2013empirical} evaluate  48 distance measures using a 1-Nearest Neighbor (1-NN) classification algorithm, on 42 time series data sets. A notable recent work is~\cite{john2020debunking}, which comprehensively evaluates 71 univariate distance measures from 5 different categories, debunking four long-standing misconceptions. For multivariate cases, Shifaz et al.~\cite{shifaz23} recently extended seven widely-used elastic measures for multivariate time series data analysis.
Unfortunately, many of the aforementioned surveys only address one or two categories of functions, lacking a systematic introduction to methods across different families for both univariate and multivariate cases.

Motivated by the aforementioned issues, this work aims to deliver an extensive review of 7 categories of state-of-the-art distance measures, covering both univariate and multivariate measures. To present each measure,  we introduce the mathematical formulas and also provide a discussion on the discrepancy and applications across various methods.

Acknowledging the fundamental distinctions between univariate and multivariate contexts, this work provides another contribution by offering a detailed comparison of the two and providing essential guidance for extending distance measures from univariate to multivariate cases, which serves as a valuable complement in this domain.  We introduce this work as a collection of the shared knowledge developed in the community on this vital task, which helps to facilitate even more successful and knowledge-guided application of distance measures in downstream tasks and the continued development of this important area of time-series data analytics.

%% file: section/S02_preliminary.tex
\section{Preliminaries and Notations}\label{sec:preliminaries}


We now introduce the formal notation for time series and distance measures. We define a univariate times-series $X=\{x_1, x_2, \dots , x_n\}$ as a sequence of real-valued numbers, where $n= |X|$ is the length of the time-series and $x_i\in \mathbb{R}$ for $i\in [1,n]$. Given this definition, we further define a multivariate, or $c$-dimensional time-series $\mathbf{X}=[\vec{X}^{(1)}, \vec{X}^{(2)}, \cdots, \vec{X}^{(c)}]$ as a set of $c$ univariate time-series of length $n$. Each row of the multivariate time series $\mathbf{X}$ represents a univariate time series $\vec{X}^{(j)}=[\vec{X}_1^{(j)}, \vec{X}_2^{(j)},\cdots,\vec{X}_n^{(j)}]$ for $j\in [1, c]$. We denote $\mathbf{X}^{(j)}_i$ as the $i$th data point on channel $j$ and $\mathbf{X}^{(j)}$ the univariate time series on channel $j$.

With the definition of univariate and multivariate time series, we can define the distance measure as follows. Take univariate time series as an example. A time-series distance measure $d$ is a function $d: \mathbb{R}^{n} \times \mathbb{R}^{n} \rightarrow \mathbb{R}$, where $d(X, Y)$ is the distance between time series $X$ and $Y$. The multivariate time-series distance measure can be defined similarly.


%% file: section/S03_taxonomy.tex
\section{Taxonomy of Time-series Distance Measure}
\label{sec:taxo}
In this section, we describe our proposed taxonomy of time-series distance measures which differentiates between measures based on 7 categories: (i) lock-step, (ii) elastic, (iii) sliding, (iv) kernel, (v) feature-based, (vi) model-based, and (vii) embedding. Figure \ref{fig:taxonomy} exhibits our proposed taxonomy and presents a visualization for each category. This taxonomy extends beyond those found in previous surveys and evaluations \cite{john2020debunking,cha,giusti2013empirical,shifaz23,ding2008querying}, not only by incorporating established categories like lock-step and elastic measures, but also by introducing two additional classes -- feature-based and model-based measures -- reflecting their widespread use in practical applications. We start by reviewing the definitions of these categories:

\begin{figure}[tb]
\centering
\includegraphics[width=\linewidth]{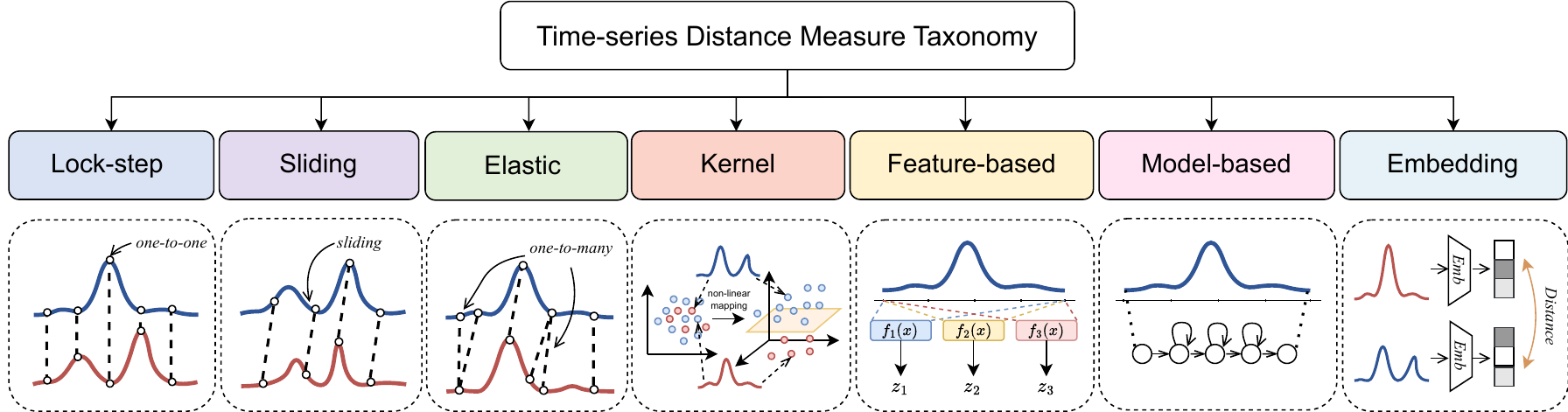}
\caption{Overview of the time-series distance measure taxonomy.}
\label{fig:taxonomy}
\vspace{-0.5cm}
\end{figure}

\noindent\textbf{Lock-step measures:} Lock-step measures assess the distance between two time series by comparing the $i$-th time step of one series with the $i$-th time step of the other, and aggregating their distances. This ``lock-step'' fashion assumes two time series are well aligned in the temporal order. The most well-known examples of the lock-step measure are the Euclidean Distance and Manhattan Distance.

\noindent\textbf{Sliding measures:}
Sliding measures calculate the lock-step distances between a time series and all shifted versions of another time series, converting the distance between the two time series into the minimum of these calculated distances. This design offers robustness to noise perturbation such as shift and translation. One of the classic examples of sliding measures is the Shape-based Distance (SBD), which serves as the distance measure for the state-of-the-art clustering algorithm known as k-Shape \cite{paparrizos2015k}.

\noindent\textbf{Elastic measures:} Elastic measures are based on the notion of temporal alignment, where the time series is first matched across its temporal range before similarity is computed. This addresses the phase alignment problem in time series data and allows the measure to conceptually ``stretch'' or ``squeeze'' the time axis to find the optimal alignment to maximize the similarity of the compared time series. This temporal elasticity is used to compare the time points in a one-to-one or one-to-many manner. In extending to the multivariate case, the alignment of time series can be approached in two ways: an ``independent'' version, where each channel is aligned separately, or a ``dependent'' version, where all channels are aligned together as a single temporal axis, accounting for their interdependencies. 

\noindent\textbf{Kernel measures:} Kernel measures employ a mapping function that projects the time series into a higher-dimensional space, before computing their distances at this space. Such measures become instrumental in some application scenarios -- such as clustering -- to project via a non-linear function to a space where the clusters are more easily separable.

\noindent\textbf{Feature-based measures:} Feature-based measures involve identifying and extracting descriptive (predominantly statistical) attributes, such as the mean value, overall trend, and other characteristics, to represent an entire time series. These attributes, also known as features, allow time series to be compared by applying simple and computationally-efficient distance measures, like Euclidean distance, to the extracted features. Descriptive features can serve as a noise-robust representation of time series, and these transformed distance measures have shown an advantage on various downstream tasks, e.g., time-series clustering \cite{christ2018time}.

\noindent\textbf{Model-based measures:} Model-based measures aim to uncover the underlying probability distribution responsible for generating the time series data, and modeling it explicitly using, e.g., the Gaussian Mixture Model (GMM) or Hidden Markov Model (HMM). 
Similar to feature-based measures, model-based measures avoid computing distance directly between raw time series data. Instead, the distance computation is performed between the learned models by utilizing straightforward measures, such as the Kullback–Leibler (KL) Divergence.

\noindent\textbf{Embedding measures:} Embedding measures generate new representations of time series in a latent space, and then compute the distances of the time series in this space, using one of the other distance measures (e.g., ED). This latent space can be used to reduce noise, increase separability, or importantly for dimensionality reduction by embedding a time series in a lower dimensional space which preserves important symmetries or properties from the original space. Unlike model-based measures, embedding measures implicitly model the distribution.
An example of an embedding used for comparison of UTS is the Generic RepresentAtIon Learning (GRAIL) framework \cite{paparrizos2019grail}.

In the following sections, we will review the 7 categories for univariate time-series distance measures (Section \ref{sec:lock-step} to Section \ref{sec:emb}), and explain how these categories can be extended to multivariate distance measures in Section~\ref{sec:multivariate}.

%% file: section/S04_lockstep.tex
\section{Lock-step Measures}\label{sec:lock-step}
Lock-step measures rely on element-wise comparison between the time series. 
Due to their relatively low cost, i.e., $\mathcal{O}(n)$ complexity where $n$ denotes the time series length, these measures have been widely applied across various fields, with the most widely used example being ED. In this section, we first review 9 well-known categories of lock-step measures~\cite{cha}. 
We also discuss three additional lock-step measures that do not belong in these 9 categories: the Dissimilarity Metric (DISSIM) \cite{frentzos}, the Autocorrelation Distance (ACD), and the Markovian Distance (MD) \cite{mirylenka}. The detailed equation for each lock-step measure can be found in Table \ref{tab:lock-step1}, Table \ref{tab:lock-step2}, and Table \ref{tab:lock-step3}.


\subsection{Minkowski-based measures}

The Euclidean distance is the classic example of a lock-step distance measure \cite{cha}. It can be calculated using the Pythagorean theorem, therefore occasionally being called the Pythagorean distance. In the late 19th century, Hermann Minkowski considered the Manhattan distance \cite{krause1986taxicab}, which has the advantage that outliers skew the result less than using the Euclidean distance. Other names for the Manhattan distance include rectilinear distance,  taxicab norm, and city block distance. 
Minkowski expanded the formulas for ED and Manhattan distance by using $p$ to denote the order of the norm, thereby generalizing these distance calculations \cite{cha}. When $p$ approaches infinity, it becomes Chebyshev distance, equivalent to finding the maximum absolute difference across all time steps. It is worth noting that, for applications with high dimensionality, lower $p$ might be more favorable; for instance, the Manhattan distance ($L_1$) is preferable to the Euclidean distance ($L_2$) in these high dimensional applications because it does not extensively penalize outliers and noise \cite{Aggarwal2001On, john2020debunking}.  The formulas for the Minkowski functions can be found in Table \ref{tab:lock-step1}.

\renewcommand{\arraystretch}{1.5}
\begin{table}[t]
\small
\caption{Lock-steps measures  (part 1), including five categories: Minkowski, $L_1$, Intersection, Inner Product, Square Chord.}
\label{tab:lock-step1}
\begin{tabular}{p{0.2\linewidth}p{0.38\linewidth}p{0.14\linewidth}}
\Xhline{2\arrayrulewidth}
\textbf{Method} & \textbf{Formula} & \textbf{Category} \\
\hline
\hline
Euclidean & $\sqrt{\sum_{i=1}^n(x_i - y_i)^2}$    & Minkowski \\ \hline
Manhattan & $\sum_{i=1}^n |x_i - y_i|$  & Minkowski \\ \hline
Minkowski & $(\sum_{i=1}^n|x_i - y_i|^p)^{\frac{1}{p}}$                   & Minkowski \\ \hline
Chebyshev & \makecell[l]{$max_i(|x_i - y_i|) = \lim_{p \rightarrow \infty} (\sum_{i=1}^n |x_i - y_i|^p)^{\frac{1}{p}}$} & Minkowski  \vspace{-0.1cm} \\ 
\hline
\hline
S{\o}rensen   & \makecell[l]{$\frac{\sum_{i=1}^n |x_i - y_i|}{\sum_{i=1}^n(x_i + y_i)}$}      & $L_1$ \\ \hline
Gower      & $\frac{1}{n} \cdot \sum_{i=1}^n |x_i - y_i|$   & $L_1$ \\ \hline

Soergel    & $\frac{\sum_{i=1}^n |x_i - y_i|}{\sum_{i=1}^n max(x_i,y_i)}$    & $L_1$ \\ \hline
Kulczynski & $\frac{\sum_{i=1}^n|x_i - y_i|}{\sum_{i=1}^n min(x_i,y_i)}$     & $L_1$ \\ \hline
Canberra   & $\sum_{i = 1}^n \frac{|x_i - y_i|}{x_i + y_i}$                 & $L_1$ \\ \hline
Lorentzian & $\sum_{i=1}^n ln(1 + |x_i - y_i|)$ & $L_1$ \\ \hline
\hline

Intersection & $\frac{\sum_{i=1}^n (|x_i - y_i|)}{2}$ & Intersection \\ \hline
Wave Hedges  & $\sum_{i=1}^n(1 - \frac{min(x_i,y_i)}{max(x_i,y_i)}) = \sum_{i=1}^n(\frac{|x_i - y_i|}{max(x_i, y_i)})$             & Intersection \\ \hline
Czekanowski  & $1 - 2\frac{\sum_{i=1}^nmin(x_i,y_i)}{\sum_{i=1}^n(x_i + y_i)}$     & Intersection \\ \hline
Motyka       & 1 - $\frac{\sum_{i=1}^nmin(x_i,y_i)}{\sum_{i=1}^n(x_i + y_i)}$     & Intersection \\ \hline
 Tanimoto  & $\frac{\sum_{i=1}^n(max(x_i,y_i) - min(x_i,y_i))}{\sum_{i=1}^nmax(x_i,y_i)}$   & Intersection \\ \hline
\hline

Inner Product & $\sum_{i=1}^n x_iy_i$ & Inner Product \\ \hline
Harmonic Mean & $2\sum_{i=1}^n (\frac{x_iy_i}{x_i + y_i})$ & Inner Product \\ \hline
Kumar-Hassebrook & $\frac{\sum_{i=1}^n(x_iy_i)}{\sum_{i=1}^n(x_i^2 + y_i^2) - \sum_{i=1}^n (x_iy_i)}$ & Inner Product \\ \hline
Jaccard & $\frac{\sum_{i=1}^n(x_i - y_i)^2}{\sum_{i=1}^n (x_i^2 + y_i^2) - \sum_{i=1}^n (x_iy_i)}$ & Inner Product \\ \hline
Cosine  & $1 - \frac{\sum_{i=1}^n x_iy_i}{\sqrt{\sum_{i=1}^nx_i^2}\sqrt{\sum_{i=1}^ny_i^2}}$ & Inner Product \\ \hline
Dice    & $\frac{\sum_{i=1}^n(x_i - y_i)^2}{\sum_{i=1}^n(x_i^2 + y_i^2)}$  & Inner Product \\ \hline
\hline

Fidelity      & $\sum_{i = 1}^n \sqrt{x_iy_i}$  & Square Chord \\ \hline
Bhattacharyya & $-ln(\sum_{i=1}^n\sqrt{x_iy_i})$      & Square Chord \\ \hline
Squared-chord & $\sum_{i=1}^n(\sqrt{x_i}-\sqrt{y_i})^2$    & Square Chord \\ \hline
  Hellinger & $\sqrt{2\sum_{i=1}^n(\sqrt{x_i}-\sqrt{y_i})^2}$ & Square Chord \\ \hline
  Matusita   & $\sqrt{\sum_{i=1}^n(\sqrt{x_i}-\sqrt{y_i})^2}$ \rule[-2ex]{0pt}{5ex} & Square Chord \\
\Xhline{2\arrayrulewidth}
\end{tabular}
\vspace{-1em}
\end{table}

\begin{table}[t]
\small
\caption{Lock-steps measures (part 2), including three categories: Squared $L_2$, Shannon's Entropy (Entropy), and Vicissitude.}
\label{tab:lock-step2}
\begin{tabular}{p{0.27\textwidth}p{0.34\textwidth}p{0.11\textwidth}}
\Xhline{2\arrayrulewidth}
\textbf{Distance Measure} & \textbf{Formula} & \textbf{Category} \\
\hline
\hline

Squared Euclidean           & $\sum_{i=1}^n (x_i - y_i)^2$ & Squared $L_2$ \rule[-2ex]{0pt}{5ex}                             \\ 
\hline
Clark                       & $\sqrt{\sum_{i=1}^n(\frac{|x_i - y_i|}{x_i + y_i})^2}$ & Squared $L_2$  \rule[-2ex]{0pt}{5ex}   \\ 
\hline
Neyman $\chi^2$                    & $\sum_{i=1}^n \frac{(x_i - y_i)^2}{x_i}$   &  Squared $L_2$  \rule[-2ex]{0pt}{5ex}           \\ 
\hline
Pearson $\chi^2$                          & $\sum_{i=1}^n \frac{(x_i - y_i)^2}{y_i}$ & Squared $L_2$ \rule[-2ex]{0pt}{5ex}               \\ \hline
Squared $\chi^2$                  & $\sum_{i=1}^n \frac{(x_i - y_i)^2}{x_i + y_i}$   & Squared $L_2$ \rule[-2ex]{0pt}{5ex}       \\ \hline
Divergence   & $2\sum_{i=1}^n\frac{(x_i - y_i)^2}{(x_i + y_i)^2}$  &  Squared $L_2$  \\ \hline
Additive Symmetric $\chi^2$      & $\sum_{i=1}^n\frac{(x_i - y_i)^2(x_i + y_i)}{x_iy_i}$ & Squared $L_2$ \rule[-2ex]{0pt}{5ex}   \\ \hline
Probabilistic Symmetric $\chi^2$    & $2\sum_{i=1}^n\frac{(x_i - y_i)^2}{x_i + y_i}$    & Squared $L_2$ \rule[-2ex]{0pt}{5ex}       \\ \hline
\hline

Kullback-Leibler  & $\sum_{i=1}^nx_iln(\frac{x_i}{y_i})$  \rule[-2ex]{0pt}{5ex}   & Entropy \\ \hline
Jeffreys          & $\sum_{i=1}^n(x_i-y_i)ln(\frac{x_i}{y_i})$ \rule[-2ex]{0pt}{5ex}        & Entropy \\ \hline
K Divergence      & $\sum_{i=1}^nx_iln(\frac{2x_i}{x_i + y_i})$ \rule[-2ex]{0pt}{5ex}      & Entropy \\ \hline
Tops{\o}e            & $\sum_{i=1}^n(x_iln(\frac{2x_i}{x_i + y_i}) + y_iln(\frac{2y_i}{y_i + x_i}))$  \rule[-2ex]{0pt}{5ex}       & Entropy \\ \hline
Jensen Shannon    & \makecell[l]{$\frac{1}{2}\sum_{i=1}^n(x_iln(\frac{2x_i}{x_i + y_i}) + y_iln(\frac{2y_i}{y_i + x_i}))$}  & Entropy \\ \hline
Jensen Difference & \makecell[l]{$\sum_{i=1}^n [\frac{x_iln(x_i) + y_iln(y_i)}{2} - \frac{x_i + y_i}{2} \cdot ln(\frac{x_i + y_i}{2})]$} & Entropy \\ 
\addlinespace[-0.2cm]
\hline
\hline


Vicis-Wave Hedges & $\sum_{i=1}^n \frac{|x_i - y_i|}{min(x_i,y_i)}$ \rule[-2ex]{0pt}{5ex} & Vicissitude \\ \hline
Emanon 2                    & $\sum_{i=1}^n \frac{(x_i - y_i)^2}{min(x_i,y_i)^2} $ \rule[-2ex]{0pt}{5ex} & Vicissitude \\ \hline
Emanon 3                    & $\sum_{i=1}^n \frac{(x_i - y_i)^2}{min(x_i,y_i)} $ \rule[-2ex]{0pt}{5ex} & Vicissitude \\ \hline
Emanon 4                    & $\sum_{i=1}^n \frac{(x_i - y_i)^2}{max(x_i,y_i)} $\rule[-2ex]{0pt}{5ex} & Vicissitude \\ \hline
Max-Symmetric $\chi^2$ & $max(\sum_{i=1}^n\frac{(x_i - y_i)^2}{x_i},\sum_{i=1}^n\frac{(x_i - y_i)^2}{y_i})$ & Vicissitude \\ \hline
\addlinespace[0.2cm]
Min-Symmetric $\chi^2$ & $min(\sum_{i=1}^n\frac{(x_i - y_i)^2}{x_i},\sum_{i=1}^n\frac{(x_i-y_i)^2}{y_i})$ & Vicissitude \\ 

\addlinespace[0.2cm]

\Xhline{2\arrayrulewidth}
\end{tabular}
\vspace{-1em}
\end{table}

\begin{table}[t]
\caption{Lock-steps measures (part 3), including 2 categories: Combination and Other functions.}\label{tab:lock-step3}
\resizebox{0.8\columnwidth}{!}{
\begin{tabular}{p{0.15\textwidth}p{0.5\textwidth}p{0.15\textwidth}}
\Xhline{2\arrayrulewidth}
\textbf{Method} & \makecell[c]{\qquad\qquad\qquad\qquad\qquad\qquad\qquad\qquad\qquad\qquad \ \ \ \ \textbf{Formula}} & \textbf{Category}\\

\hline
\hline

Taneja                  & {\begin{align*}\sum_{i=1}^n\frac{(x_i + y_i)}{2} \cdot \ln(\frac{x_i + y_i}{2\sqrt{x_iy_i}})\end{align*}} \rule[-2ex]{0pt}{5ex} & Combination\\ 
\hline
Kumar-Johnson           & {\begin{align*}\sum_{i=1}^n\frac{(x_i^2 - y_i^2)^2}{2(x_iy_i)^{\frac{3}{2}}}\end{align*}}  \rule[-2ex]{0pt}{5ex}  & Combination     \\ 
\addlinespace[-0.3cm]
\hline
\addlinespace[0.2cm]
Avg($L_1$,$L_{\infty}$) & {\begin{align*}\frac{\sum_{i=1}^n(|x_i - y_i|) + max_{i}|x_i - y_i|}{2} \end{align*}}   & Combination    \\
\addlinespace[0.2cm]
\hline
\hline

\addlinespace[-1cm]
\makecell[l]{\\ \\ \\ \\ \\ \\DISSIM} & {\begin{align*}
    DISSIM(X, Y) &= \sum_{k=1}^{n-1}\int_{t_k}^{t_{k+1}}D_{X, Y}(t)dt \\
    D_{X, Y}(t) &= \sqrt{at^2+bt+c}
\end{align*}}
& \makecell[l]{\\ \\ \\ \\ \\ \\Other}
\\
\addlinespace[-0.3cm]
\hline

\addlinespace[0.2cm]
PCC & {\begin{align*}
    \frac{\sum_{i=1}^{n}(x_i-\mu_{X})(y_i-\mu_{Y})}{\sqrt{\sum_{i=1}^n(x_i-\mu_X)^2}\sqrt{\sum_{i=1}^n(y_i-\mu_Y)^2}}
\end{align*}
} & Other \\
\addlinespace[0.2cm]
\hline
\addlinespace[-1cm]

\makecell[l]{\\ \\ \\ \\ \\ACD} & 
{
\begin{align*}
R &= \{r(\tau)\}_{\tau=1}^{\tau=n} \\ 
r(\tau) &= \frac{E[(x_t-\mu)(x_{t+\tau}-\mu)]}{\sigma^2} \\
\end{align*}
} & \makecell[l]{\\ \\ \\ \\ \\ Other} \\
\addlinespace[-0.5cm]
\hline
\addlinespace[-2cm]
\makecell[l]{ \\ \\ \\ \\ \\ \\ \\ \\ \\MD} & {\begin{align*}
    M(x_{t-k}, x_{t-k+1}, ..., x_{t}) &= Pr[x_t|x_{t-1}, ..., x_{t-k}] \\
    &= \frac{Freq[x_t, x_{t-1}, ..., x_{t-k}]}{Freq[x_{t-1}, ..., x_{t-k}]} \\
    Pr(y_1, ..., y_n|M) &= Pr[y_1, ..., y_k]\Pi_{t = k+1}^{n}M(y_{t-k}, ..., y_t)
\end{align*}
\begin{align*}
MD\ Distance = logPr(y_1, ..., y_n|M) \sim \sum_{t=k+1}^{n}log[M(y_{t-k}, y_{t-k+1}, ..., y_t)]
\end{align*}}  & \makecell[l]{\\ \\ \\ \\ \\ \\ \\ \\ Other} \\
\addlinespace[-0.2cm]
\Xhline{2\arrayrulewidth}
\end{tabular}}
\vspace{-1em}
\end{table}

\renewcommand{\arraystretch}{1}

\subsection{$L_1$ Functions}

The $L_1$ functions all involve adapted versions of the Manhattan metric. S{\o}rensen distance \cite{sorensen} is a normalized adaptation of the $L_1$ distance that confines its values to the range $[0,1]$. It is widely used in the fields of ecology and environmental sciences~\cite{looman}. 
Similar to S{\o}rensen distance, Gower distance~\cite{gower} also normalizes the Manhattan distance, but in this case by the length of the time series. 
It has been shown to be effective for mixed continuous and categorical variables \cite{tuerhong2014gower}. 
The Soergel distance \cite{monev} normalizes the $L_1$ distance using the sum of the maximum values of the corresponding elements in the two compared time series, whereas the Kulczynski distance \cite{deza} normalizes by the sum of the minimum values of respective elements from the two time series. Canberra distance has a strong sensitivity to small changes near zero, so Canberra distance is often used for data scattered around an origin \cite{gordon}. Lorentzian distance \cite{deza} applies a logarithm operation to the $L_1$ distance, and the constant term is added to avoid $\log(0)$ issues. Compared with other $L_1$ measures, Lorentzian has shown robustness to noise and outliers. 
In prior evaluation studies, it has been shown that Lorentzian distance, when applied with normalization strategies such as z-score, significantly outperforms Euclidean Distance in the classification downstream task \cite{john2020debunking}. The formulas can be found in Table \ref{tab:lock-step1}.

\subsection{Intersection Functions}
The intersection family of functions shares a strong connection with the $L_1$ family. Although there exist some exceptions, many similarity measures within the intersection family can be converted into the distance measures in the $L_1$ family by using the formula  \cite{cha}, i.e., $d(X, Y) = 1 - s(X, Y)$, where $d$ and $s$ denote the dissimilarity and similarity measure between two time series. The Czekanowski distance, derived from Czekanowski similarity, $s_{Cze}=2\frac{\sum_{i=1}^nmin(x_i,y_i)}{\sum_{i=1}^n(x_i + y_i)}$, is an application of this transformation rule. This measure is equivalent to S{\o}rensen distance in the $L_1$ family through a mathematical formula transformation. Another transformation example could also be found in the Tanimoto distance and its corresponding member of the $L_1$ family, the Soergel distance. In prior evaluation studies \cite{giusti2013empirical, john2020debunking}, the intersection family functions are not able to surpass the basic Euclidean distance in downstream applications. However, there is still value in these distances for specific tasks and circumstances \cite{duda,hedges}. The equations for measures in this category can be found in Table \ref{tab:lock-step1}.

\subsection{Inner Product Functions}
This family of methods incorporates the inner product for similarity measures, which is the sum of the element-wise multiplication of two vectors or time series (with proper transformation, the functions can be applied to measure dissimilarity). 
From a geometric perspective, the inner product boils down to the ratio of magnitudes between a vector, and the projection of another vector onto that first vector.
By normalizing these vectors' $L_2$-norms, this function effectively captures the angular information in space between two time series, e.g., cosine similarity. 
The inner product also has a direct connection to Pearson correlation, which is a measure of linear dependence between time series.
Specifically, Pearson correlation boils down to the inner product of two z-normalized time series; where the normalization ensures that the time series are of the same scale and variance, and the inner product then measures the linear dependence between the two time series (i.e., vectors).
This connection is further demonstrated in the following paragraph.
Harmonic Mean similarity \cite{deza} calculates element-wise harmonic means between the time series. 
It is often used when focusing on rates of change. 
Kumar-Hassebrook \cite{kumar} is similar to harmonic mean distance but measures the Peak-to-correlation energy; a frequently used algorithm for comparing patterns in signals from digital image sensors. Jaccard \cite{jaccard} and Dice distance \cite{dice} are widely used in research fields such as information retrieval \cite{dice}. Table \ref{tab:lock-step1} displays formulas for measures that belong to this category. 

As mentioned above, inner product functions have shown strong relationships with other families. 
Here, we examine the relationship between inner product, ED, Pearson correlation coefficient (PCC, will be discussed later in more detail), and Cosine similarity, and discuss how these lead to equivalence of different problems, under certain conditions.  Let us first consider Inner product and Squared ED:

\begin{align*}
    \text{Inner\_Product} (X,Y) &= \sum_i^{n} x_i y_i = X^\top Y \\
    \text{Squared\_ED} (X,Y) &= \sum_i^{n} (x_i-y_i)^2 \\ 
        &= \norm{X}_2^2 + \norm{Y}_2^2 - 2X^\top Y.
\end{align*}

When time series $X$ and $Y$ have unit lengths, the squared ED becomes $2(1-X^\top Y)= 2(1-\text{Inner\_Product} (X,Y))$. This means that, if $\norm{X}_2=\norm{Y}_2=1$, the nearest-neighbor search (NSS) problem is equivalent to the problem of finding the pair of time series that maximizes the inner product, i.e., the maximum inner product search problem~\cite{shrivastava}. 
The Pearson correlation coefficient (PCC) also has a relationship with inner product, as follows:

\begin{align*}
    \text{PCC} (X,Y) &= \frac{\sum_{i=1}^{n}(x_i-\mu_{X})(y_i-\mu_{Y})}{\sqrt{\sum_{i=1}^n(x_i-\mu_X)^2}\sqrt{\sum_{i=1}^n(y_i-\mu_Y)^2}} \\
    &= \frac{X^\top Y}{\norm{X}\norm{Y}} \text{, \ \  if X, Y are zero-centered} \\
    &= X^\top Y \text{, \ \ if X, Y are zero-centered, with unit lengths}
\end{align*}
 where $\mu_X$ and $\mu_Y$ denote the mean value of two time series.
Finally, Cosine similarity $Cosine(X,Y) = \cos(\theta) = \frac{ X^\top Y}{\norm{X}_2 \norm{Y}_2 }$ equals to $InnerProduct(X,Y)$ when $\norm{X}_2=\norm{Y}_2=1$.

\subsection{Squared Chord Functions}
The squared chord functions are a collection of measures that incorporate the sum of geometric means of two time series, which is computed by summing the square roots of the products of corresponding elements from two time series. Fidelity similarity \cite{deza} is built by the sum of geometric means given two time series. Bhattacharyya distance \cite{bhattacharyya} uses a statistical distance measure that measures the dissimilarity of two probability distributions, which is the general case of Mahalanobis distance. Squared chord distance \cite{gavin}, Matusita distance \cite{matusita}, and Hellinger distance\cite{deza} are similar measures that are capable of emphasizing more dissimilar features. Squared Chord functions are widely applied in biological data analysis, e.g., pollen records, where they demonstrate superior performance in statistical analysis compared to other distance measures \cite{gavin2003statistical}. Table \ref{tab:lock-step1} exhibits equations for measures in this category.

\subsection{Squared $L_2$ Functions}



The squared $L_2$ functions, or $\chi^2$ functions, are a group of distance measures that have the squared Euclidean distance as the dividend. Squared Euclidean distance takes basic form without normalization. Clark distance \cite{deza}  normalizes the Euclidean distance with the sum of element pairs from two time series. Neyman $\chi^2$ \cite{neyman} and Pearson $\chi^2$ divergence \cite{pearson} derive their denominators from either the first or the second of the two time series under comparison. As these two formulas are asymmetric, they are categorized as divergence functions. To address this issue, several symmetric versions have been proposed. Probabilistic Symmetric $\chi^2$ distance \cite{deza}, Squared $\chi^2$ distance \cite{gavin}, and Divergence distance \cite{cox} compute the sum of the element pairs as the denominator. This can be considered a symmetric version of the Neyman $\chi^2$ distance. To the same end, Additive Symmetric $\chi^2$ distance \cite{deza} computes the product of the element pairs as the denominator. Among this family, measures like Clark have been widely used in various downstream tasks such as light curve classification in astronomy \cite{chaini2024light}. Previous evaluation studies on K-Nearest Neighbor (KNN) classifier \cite{abu2019effects}
, have demonstrated that the symmetric structures within the Squared $L_2$ family, including the Additive Symmetric $\chi^2$ distance and Probabilistic Symmetric $\chi^2$, do not inherently guarantee superior performance compared to asymmetric divergence. Among them, Squared $\chi^2$ and Clark Distance have shown supreme performance compared to others. Table \ref{tab:lock-step2} displays formulas for measures that belong to the squared $L_2$ category.

\subsection{Shannon's Entropy Functions}



The following functions are based on Shannon's Entropy measure which has to deal with how much information a variable contains and the probabilistic uncertainty of information. Kullback-Leibler (KL) divergence \cite{kullback}, also relative entropy, has been widely adopted for capturing how one probability distribution diverges from the other. However, its asymmetrical nature may pose challenges in specific applications where a metric and its associated properties are needed. To solve this problem, Jeffreys distance \cite{kullback, jeffrey,taneja} is considered to be the symmetric version of Kullback-Leibler distance. Jensen-Shannon distance and Tops{\o}e distance \cite{deza} are symmetric versions of K divergence distance, which shows a structure similar to KL divergence. Tops{\o}e distance is also referred to as $information$ $statistics$ \cite{gavin}. Investigating the concept of information radius, which emerges from the concavity of Shannon's entropy, Sibson \cite{sibson} introduced the equation of Jensen difference \cite{cha}. For its entropy nature, functions in Shannon's Entropy family are widely used in numerous downstream tasks such as time series classification, clustering, and anomaly detection \cite{wu2019learning,barz2018detecting,lee2014clustering}. Table \ref{tab:lock-step2} displays formulas for measures in this category.

\subsection{Vicissitude Functions}

This group of functions is based on Vicis-Wave Hedges function \cite{cha}. Vicis-Wave Hedges distance, also named Emanon 1 distance, is a variant of the Wave Hedges function. It is built by applying the relationship between Sorensen and Canberra to Kulczynski \cite{cha}. Emanon 2 and 3 distance \cite{cha} are variants of Vicis-Wave Hedges where the squared operation is applied for the numerator and denominator. Emanon 4 distance \cite{cha} takes the sum of the squared difference over the maximum of the element pairs. Max-Symmetric $\chi^2$ distance \cite{cha} takes the maximum of the Pearson and Neyman $\chi^2$, while Min-Symmetric $\chi^2$ distance \cite{cha} takes the minimum. Prior evaluation studies \cite{john2020debunking} demonstrate that Emanon 4 outperforms other functions within the Vicissitude family, offering significantly better performance than ED in the classification task. Table \ref{tab:lock-step2} displays formulas for measures that belong to this category.

\subsection{Combination Functions}


The combination functions integrate techniques from various aforementioned families. For example, Taneja measure \cite{taneja} is also known as arithmetic-geometric mean divergence measure. Kumar-Johnson \cite{kumar2005symmetric} combines strategies from symmetric $\chi^2$, arithmetic-geometric mean divergence. Avg($L_1$,$L_\infty)$ is the average between the $L_1$ distance and Chebyshev distance. By seizing elements from different methods, combination functions can leverage the diverse benefits each family offers, potentially yielding more versatile and robust measures under different scenarios. In prior evaluation studies \cite{john2020debunking,giusti2013empirical}, Avg($L_1$,$L_\infty)$ shows superior classification performance compared with Taneja and Kumar-Johnson. Formulas for measures mentioned above are in Table \ref{tab:lock-step3}.

\subsection{Other Functions}

In this section, we present three other measures that show differences from the previously mentioned families. These metrics offer more flexible handling of time.

Conventional Lockstep Measures are not designed to address variations in the temporal dimension of time series, typically assuming uniform lengths and sampling rates across time steps. Dissimilarity Metric (DISSIM) \cite{frentzos} accounts for these issues by calculating the definite integral of the function of time of the ED between two time series during each time interval and sums up the distance in all intervals. Formally, DISSIM \cite{frentzos} is defined in Table \ref{tab:lock-step3}, where $[t_1, t_n]$ is the time period, $D_{X, Y}$ is the Euclidean distance between two points moving with linear functions of time between consecutive timestamps, and $a, b, c$ $(a\geq 0)$ are factors of the trinomial. In prior evaluation studies, it has shown to be effective on various downstream tasks such as classification and clustering \cite{john2020debunking,kotsakos2018time}


Pearson correlation coefficient (PCC) is a widely used similarity measure to capture the linear correlation between two data samples (with proper transformation it can be applied to quantify the distance). The formula is shown in Table \ref{tab:lock-step3}, where $\mu_X$ and $\mu_Y$ denote the mean value of each data. It has enlightened the design of many following correlation-based functions. Autocorrelation Distance (ACD) \cite{mirylenka} calculates the autocorrelation vector that consists of autocorrelation coefficients with different lags, where the number of coefficients included is a parameter to tune. ACD distance is then defined by calculating computing the Euclidean distance between autocorrelation vectors. The autocorrelation vector, autocorrelation coefficients \cite{mirylenka} are defined in Table \ref{tab:lock-step3}, where $\mu$ and $\sigma^2$ are the mean and variance of the time series. Research has shown it to be robust against both stationary and non-stationary time series \cite{riyadi2017clustering}.

    

Markovian Distance (MD) \cite{mirylenka} defines the similarity between two time series $X$ and $Y$ as the probability that $Y$ is generated using a Markov model characterized by $X$. MD computes such probability by first estimating a transition probability matrix $M$ that characterizes a Markov Chain by estimating the conditional probabilities of query $X$. This is achieved by calculating frequencies of all sequences of length $k$ and $k+1$, where $k$ (a parameter to be estimated) is the number of previous states the current state depends on. The distance between $X$ and $Y$ is then calculated as the probability of generating $Y$ using the model of $X$.
Since there is only one query series $X$, it is hard to estimate the initial states, which are instead set to be equitable at first. Entries in the transition matrix with the same prefix $[x_{t-1}, ..., x_{t-k}]$ form a probability distribution, and the ``equally-shared probability" of prefixes not observed in the query will be divided among other observed prefixes. Logarithmic operation is used to avoid the accumulation of machine errors. 
The Markovian Distance \cite{mirylenka} is formally defined in Table \ref{tab:lock-step3}. In real applications, both ACD and MD have been widely utilized to detect dissimilarities between data streams  \cite{alseghayer2019dcs}.

\begin{figure}[t]
\centering
\includegraphics[width=0.9\linewidth]{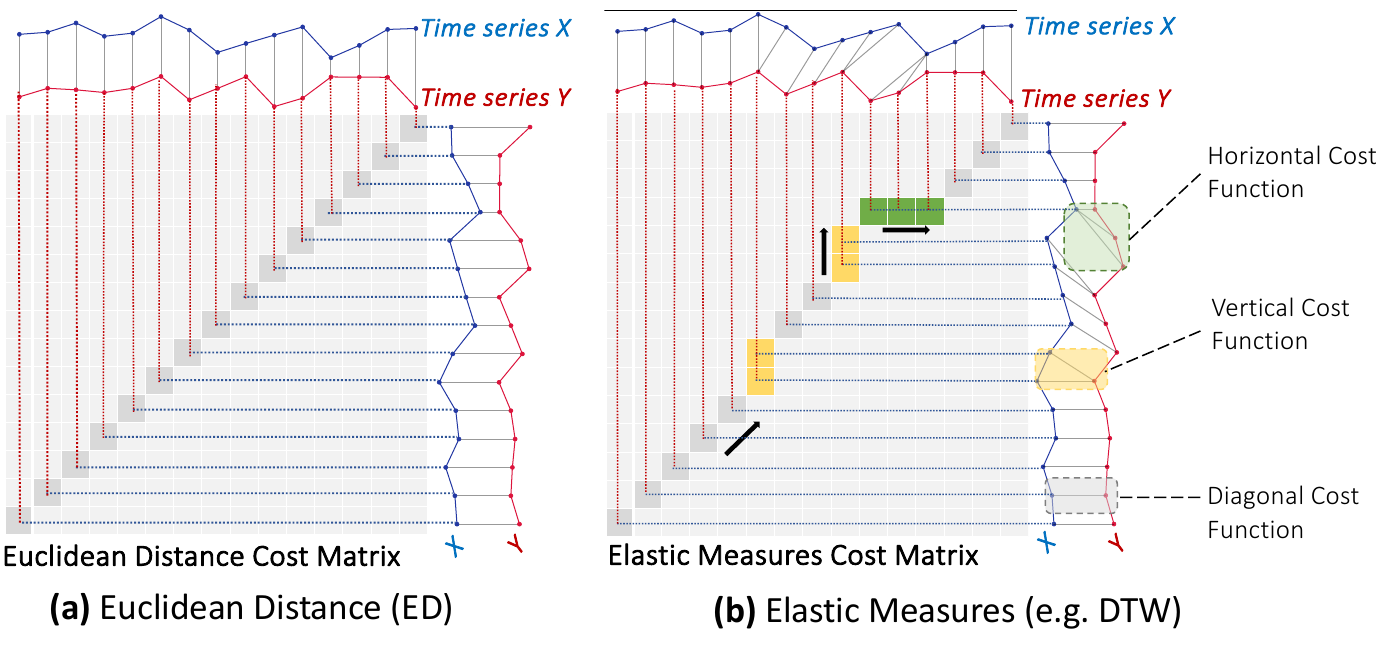} \\
\caption{Alignments and cost matrices of Euclidean Distance (ED) and Dynamic Time Warping (DTW)}
\label{fig:EDvsDTW}
\vspace{-0.3cm}
\end{figure}

%% file: section/S05_elastic.tex
\section{Elastic Measures}
\label{sec:elastic}
For time series data with phase misalignments, stretching or squeezing of observations over the time range, or fluctuations, lock-step comparisons are not always a suitable distance measure due to the inability to consider these distortions \cite{bagnall2017great, ding2008querying, john2020debunking, wang2013experimental}. Figure \ref{fig:EDvsDTW} illustrates the linear mapping of ED and the one-to-many alignment of elastic measures such as Dynamic Time Warping (DTW) that captures the shape similarity 
of two time series.

Among dozens of distance measures proposed to align such distortions, \emph{elastic measures}, which create a non-linear mapping between time series to align or stretch their points, have shown to be effective in numerous downstream tasks such as classification and clustering. In addition, contrary to prior beliefs, ED may not converge to the high accuracy of elastic measures with increasing dataset sizes \cite{john2020debunking}. 

One limitation of elastic measures is their quadratic time complexity (i.e., $\mathcal{O}(n^2)$, for time series of length $n$), whereas the lock-step measures like ED have linear time complexity (i.e., $\mathcal{O}(n)$). In large-scale settings, the higher complexity of elastic measures  results in a runtime overhead often between one to three orders of magnitude over ED \cite{john2020debunking,bagnall2017great,tan2020fastee}, which usually prevents applications from using elastic measures and rely instead on less accurate measures. Consequently, several acceleration methods, such as {\em lower bounding} and {\em early abandoning}, are developed to speedup the application of elastic measures in tasks like the K-NN search in large-scale settings.


This section reviews elastic measures by starting with \emph{Dynamic Time Warping} (DTW), the earliest and most popular elastic measure, and provides a generalized formula to showcase the recursive (dynamic programming) computation shared by all elastic measures and highlights the different cost functions across elastic measures. Based on these different cost functions, elastic measures developed after DTW are categorized into {\em threshold-based} elastic measures and {\em metric} elastic measures \cite{paparrizos2023accelerating}. Acceleration methods of elastic measurs, including {\em lower bounding} and {\em early abandoning}, are reviewed at the end of this section.


\subsection{Dynamic Time Warping (DTW)} 
 Dynamic Time Warping (DTW) addresses distortions or phase differences by permitting one-to-many point matching to achieve local alignment, so the two time series are aligned based on their optimal shape similarity.  To find the local alignment, DTW finds an optimal alignment path and the minimum distance between two time series by computing a distance matrix, $D$, using the following recursive computation: 
\vspace{-0.1cm}
\begin{equation}
\label{eqn:dtw_recursion}
\footnotesize{
    D(i,j) = \left.
    \begin{cases}
        (x_i-y_j)^2 & \text{if $i,j = 1$}\\
        D(i-1,j) + (x_i-y_j)^2 & \text{if $i \neq 1$ and $j = 1$} \\
        D(i,j-1) + (x_i-y_j)^2 & \text{if $i = 1$ and $j \neq 1$} \\
        min \begin{cases}D(i-1,j-1) + (x_i-y_j)^2 \\ D(i-1,j) + (x_i-y_j)^2 \\
        D(i,j-1) +(x_i-y_j)^2 \end{cases} & \text{if $i, j \neq 1$}
    \end{cases}
    \right.
}
\vspace{-0.1cm}
\end{equation}

\noindent The optimal alignment path, $W = \{w_1, w_2, ..., w_p\}$, which starts from the bottom-left corner and ends at the top-right corner in the matrix where the distance of alignments add up to the cell in the top-right corner: $$DTW(X, Y) = D(n_X, n_Y) = \sum_{i=1}^{p}(x_{{w_i}[1]} - y_{{w_i}[2]})^2$$.  

\noindent The warping path follows two properties \cite{tan2019elastic}:
\begin{itemize}[noitemsep,topsep=0pt]
    \item \textbf{Boundary Constraints}: $w_1 = (1, 1)$ and $w_p = (n_X, n_Y)$, meaning the optimal warping path starts on the bottom-left corner of $D$ and ends on the upper-right corner of $D$.
    
    \item \textbf{Continuity and Monotonicity}: if $w_i = (i, j)$ for $i \in [2, p-1]$, then $w_{i+1} \in \{(i+1, j), (i, j+1), (i+1, j+1)\}$, meaning that the warping path, starting from bottom-left, only moves vertically upwards, horizontally towards the right, or diagonally towards top-right continuously until arriving at the top-right corner. DTW uses the same distance function (i.e., squared difference) in each matrix cell regardless of whether the optimal path arrives at that cell horizontally, vertically, or diagonally. 
\end{itemize}
Several extensions are developed for DTW to enhance its performance in terms of speed and/or accuracy, and these extensions are applicable to other elastic measures as they share the same dynamic programming structure with DTW. DTW variants with such extensions include: \begin{itemize}
    \item \textbf{Constrained DTW: } Locality constraints are commonly applied to DTW to reduce runtime and improve classification accuracy \cite{john2020debunking}; locality constraints limit the range of warping allowed to avoid unreasonably far-reaching alignments. This approach is referred to as Constrained DTW (cDTW) \cite{sakoe1978dynamic}, and the most widely adopted locality constraint is the Sakoe-Chiba band \cite{sakoe1978dynamic}. Commonly referred to as the \emph{warping window}, the Sakoe-Chiba band is mathematically defined as the maximum possible deviation of the alignment path from the diagonal of $D$, and cells further away are not computed.  

    \item \textbf{Weighted DTW: } Weighted DTW (WDTW) \cite{jeong2011weighted} creates a weighted vector that penalizes the differences between $i$ and $j$ and thereby better captures the shape similarity of two time series. WDTW computes weights by adjusting the parameters of a logistic function and is capable of giving linear weights, sigmoid weights, two distinct weights, or constant weights to alignments in the warping path. 

    \item \textbf{Derivative DTW: } Derivative DTW \cite{keogh2001derivative} improves DTW's ability to capture shape similarities by replacing each element of the original two time series with a ``derivative'' value that measures the changes in the shape of the time series at that time step. For noisy datasets, exponential smoothing can be applied before computing the "derivative" to further improve shape similarity capture. 
\end{itemize}

\begin{table}[t!]
\centering
    \normalsize
    \caption{Summary of distances ($dist^D(x_i, y_j)$, $dist^H(x_i, y_j)$, and $dist^V(x_i, y_j)$), transformation functions ($trans(D(n_{X}, n_{Y}))$) for threshold-based elastic distances. \cite{paparrizos2023accelerating}}
\label{table:threhold-based-measures}
    \vspace{-0.3cm}
        \resizebox{0.55\linewidth}{!}{
        \begin{tabular}{|l|l|l|}
            \hline
            \multirow{4}{*}{ LCSS \cite{vlachos2002discovering} } & $dist^D(x_i, y_j)$ & {$\begin{cases} 
                                            1 \textup{ if } |x_i - y_j| \leq \epsilon \\
                                            0 \textup{ otherwise }
                                        \end{cases}$} \rule[-2ex]{0pt}{5ex}  \\\cline{2-3}
                                 & $dist^V(x_i, y_j)$ & 0 \rule[-2ex]{0pt}{5ex}  \\\cline{2-3}
                                 & $dist^H(x_i, y_j)$ & 0 \rule[-2ex]{0pt}{5ex}  \\\cline{2-3}
                                 & $trans(D(n_{X}, n_{Y}))$ & $1 - \frac{D_(n_{X}, n_{Y})}{min(n_{X}, n_{Y})}$ \rule[-2ex]{0pt}{5ex} \\\cline{2-3}
            
            \hline
            \multirow{4}{*}{ EDR \cite{chen2005robust}} & $dist^D(x_i, y_j)$ & $\begin{cases} 
                                            0 \textup{ if } |x_i - y_j| \leq \epsilon \\
                                            1 \textup{ otherwise }
                                        \end{cases}$ \\\cline{2-3}
                                 & $dist^V(x_i, y_j)$ & 1 \rule[-2ex]{0pt}{5ex}  \\\cline{2-3}
                                 & $dist^H(x_i, y_j)$ & 1 \rule[-2ex]{0pt}{5ex}  \\\cline{2-3}
                                 & $trans(D(n_{X}, n_{Y}))$ & $D(n_{X}, n_{Y})$ \rule[-2ex]{0pt}{5ex}  \\\cline{2-3}
            \hline
            \multirow{4}{*}{   SWALE \cite{ding2008querying}} & $dist^D(x_i, y_j)$ & $\begin{cases} 
                                            r \textup{ if } |x_i - y_j| \leq \epsilon \\
                                            p \textup{ otherwise }
                                        \end{cases}$ \\\cline{2-3}
                                 & $dist^V(x_i, y_j)$ & p \rule[-2ex]{0pt}{5ex}  \\\cline{2-3}
                                 & $dist^H(x_i, y_j)$ & p \rule[-2ex]{0pt}{5ex}  \\\cline{2-3}
                                 & $trans(D(n_{X}, n_{Y}))$ & $D(n_{X}, n_{Y})$ \rule[-2ex]{0pt}{5ex}  \\ \cline{2-3}
            \hline
        \end{tabular}}
        \vspace{-0.3cm}
\end{table}

    

\begin{table}[t!]
    \caption{Summary of distances ($dist^D(x_i, y_j)$, $dist^H(x_i, y_j)$, and $dist^V(x_i, y_j)$), transformation functions ($trans(D(n_{X}, n_{Y}))$) for metric elastic distances.}
     \vspace{-0.3cm}
\label{table:metric}
        \resizebox{0.85\linewidth}{!}{
        \begin{tabular}{|l|l|l|}

    \hline
        \multirow{5}{*}{ERP \cite{chen2004marriage}} & $dist^D(x_i, y_j)$ & $(x_i - y_j)^2$ \rule[-2ex]{0pt}{5ex}  \\\cline{2-3}
                             & $dist^V(x_i, y_j)$ & $(x_i - g)^2$ \rule[-2ex]{0pt}{5ex}  \\\cline{2-3}
                             & $dist^H(x_i, y_j)$ & $(y_j - g)^2$ \rule[-2ex]{0pt}{5ex}  \\\cline{2-3}
                             & $trans(D(n_{X}, n_{Y}))$ & $D(n_{X}, n_{Y})$ \rule[-2ex]{0pt}{5ex}\\ \cline{2-3}
                            
        \hline
        \multirow{5}{*}{ MSM \cite{stefan2013move}} & $dist^D(x_i, y_j)$ & $|x_i - y_i|$ \rule[-2ex]{0pt}{5ex}  \\\cline{2-3}
                             & $dist^V(x_i, y_j)$ & $\begin{cases}
                                    c \textup{ if } x_{i-1} \leq x_i \leq y_j \textup{ or } x_{i-1}  \geq x_i \geq y_j \\
                                    c + min \begin{cases}
                                                |x_i - x_{i-1}| \\
                                                |x_i - y_j|
                                            \end{cases} \textup{ otherwise}
                                    \end{cases}$ \rule[-2ex]{0pt}{5ex}  \\\cline{2-3}
                             & $dist^H(x_i, y_j)$ & $\begin{cases}
                                    c \textup{ if } y_{j-1} \leq y_j \leq x_i \textup{ or } y_{j-1} \geq y_j \geq x_i \\
                                    c + min \begin{cases}
                                                |y_j - y_{j-1}| \\
                                                |y_j - x_i|
                                            \end{cases} \textup{ otherwise}
                                    \end{cases}$ \rule[-2ex]{0pt}{5ex}  \\\cline{2-3}
                             & $trans(D(n_{X}, n_{Y}))$ & $D(n_{X}, n_{Y})$ \rule[-2ex]{0pt}{5ex}  \\\cline{2-3}
        \hline
        \multirow{5}{*}{ TWED \cite{marteau2008time}} & $dist^D(x_i, y_j)$ & $(x_i-y_j)^2+ (x_{i-1}- y_{j-1})^2
                                        + \nu(|t_{x_i} - t_{x_{i-1}}| + |t_{y_j} - t_{y_{j-1}}|)$\rule[-2ex]{0pt}{5ex}   \\\cline{2-3}
                             & $dist^V(x_i, y_j)$ & $(x_i - x_{i-1})^2 + \nu(|t_{x_i} - t_{x_{i-1}}|) + \lambda$\rule[-2ex]{0pt}{5ex}   \\\cline{2-3}
                             & $dist^H(x_i, y_j)$ & $(y_j - y_{j-1})^2 + \nu(|t_{y_j} - t_{y_{j-1}}|) + \lambda$ \rule[-2ex]{0pt}{5ex}  \\\cline{2-3}
                             & $trans(D(n_{X}, n_{Y}))$ & $D(n_{X}, n_{Y})$ \rule[-2ex]{0pt}{5ex}  \\\cline{2-3}
                             
        \hline
        \end{tabular}
        }
        \vspace{0.2cm}
\end{table}

\subsection{Threshold-based and Metric Elastic Measures}

Numerous elastic measures are proposed after DTW to overcome its limitations such as not being a metric or poor performance on noisy datasets. To these ends, subsequent elastic measures adopt different cost functions for diagonal and vertical/horizontal movements (whereas DTW uses the same squared difference cost function for all movements) and sometimes introduce additional parameters; however, they all use a dynamic programming approach to find the optimal warping path, which could be generalized as: 
\vspace{-0.1cm}
\begin{equation}
\footnotesize{
    D(i,j) = \left.
    \begin{cases}
        initial\_distance(x_i,y_j) \qquad\qquad\qquad\quad \text{if $i,j = 1$}\\
        D(i-1,j) + dist^V(x_i,y_j) \qquad\quad\;\: \text{if $i \neq 1$ and $j = 1$} \\
        D(i,j-1)+ dist^{H}(x_i,y_j) \qquad\quad\;\: \text{if $i = 1$ and $j \neq 1$} \\
        min \begin{cases}D(i-1,j-1) + dist^D(x_i,y_j) \\ D(i-1,j) + dist^{V}(x_i,y_j) \\
        D(i,j-1) + dist^{H}(x_i,y_j) \end{cases} \    \text{if $i, j \neq 1$}
    \end{cases}
    \right.
}
\vspace{-0.1cm}
\label{equation: generalized elastic measure}
\end{equation}
where $dist^D(x_i, y_j)$, $dist^{V}(x_i, y_j)$, and $dist^{H}(x_i, y_j)$ are the cost functions for diagonal, vertical, and horizontal movements, respectively, and $initial\_distance(x_i, y_j)$ is the cost function for initial alignment in $D(1, 1)$. Based on different types of cost functions, elastic measures are categorized into \emph{Threshold-based} and \emph{metric} elastic measures. 

\textbf{Threshold-based Elastic Measures}: To improve DTW's ability to handle outliers in noisy datasets, threshold-based elastic measures use a threshold parameter $\epsilon$ to decide whether two elements match or not; such binary classification of the relationship between two elements from two time series regardless of their numerical difference makes threshold-based elastic measures robust against outliers. In computing the cost matrix, a match and mismatch correspond to a diagonal cost or horizontal/vertical cost respectively. Table \ref{table:threhold-based-measures} summarizes distance functions and transformation functions, for three popular threshold-based elastic distance measures.

\begin{itemize}[noitemsep,topsep=0pt]
    \item \textbf{Longest Common Subsequence (LCSS)} \cite{vlachos2002discovering}, initially developed for pattern matching within text strings, has been adapted to evaluate similarity between time series data. In this context, LCSS increases the similarity score by 1 for each matching and by 0 for each mismatch. The resulting LCSS distance measures the \emph{similarity} of the two time series, and the transformation function is applied to the upper-corner cell in the diagonal matrix to convert the similarity score to a distance measure. 
    
    \item \textbf{Edit Distance on Real Sequences (EDR)} \cite{chen2005robust} is an adaptation of edit distance for strings to time series distance measure. EDR achieves robustness against outliers by quantizing the distance between elements to either 0 or 1, thus reducing the impact of outliers. 
    
    \item \textbf{Sequence Weighted Alignment (SWALE)} \cite{ding2008querying} generalizes EDR by incorporating a parameter $r$ for a match and a penalty parameter $p$ for a mismatch, instead of fixed 1 and 0 as in EDR. 
\end{itemize}

\textbf{Metric Elastic Measures}: DTW and threshold-based measures are not metric distances and cannot use the triangle inequality \cite{stefan2013move} to take advantage of generic indexing methods \cite{hjaltason2003index, yianilos1993data, samet2003properties}, clustering methods \cite{brisaboa2008clustering, indyk1999asublinear, ganti1999clustering}, and pruning methods \cite{chen2004marriage} designed for metric distances. Therefore, several metric elastic measures with different cost functions were proposed; as summarized in Table \ref{table:metric}, there are three popular metric elastic measures: 
\begin{itemize}[noitemsep,topsep=0pt] 
    \item \textbf{Edit Distance with Real Penalty (ERP)} \cite{chen2004marriage} introduces an additional gap value parameter, $g$, to compute the distance for horizontal and vertical movements and uses the squared difference ($(x_i - y_j)^2$) to compute the diagonal movements. Such design makes ERP a metric but it suffers from the inability to handle vertically shifted time series.
    
    \item  \textbf{Move-Split-Merge (MSM)}\cite{stefan2013move} aligns two time series using $move$ (diagonal movement along the cost matrix),  $split$ (replicate and stretch the previous element), and $merge$ (merging identical values into a single element) operations, which allows MSM to become a translation-invariant metric distance. 
    
    \item \textbf{Time Warp Edit Distance (TWED)} \cite{marteau2008time} imposes a penalty on the differences in timestamps in addition to the numerical differences and also introduces a stiffness parameter $\lambda$ to restrict far-reaching alignments. 
    
\end{itemize}

\subsection{Acceleration Methods for Elastic Measures}
The most common methods for accelerating elastic measures in applications such as classification involves the use of {\em lower bounding} or {\em early abandoning}. 

\textbf{Early Abandoning (EA):} In nearest neighbor search tasks, EA \cite{trillions_2012,herrmann} monitors the accumulated distance during computation of a full elastic measure distance and abandons the computation when the accumulated distance is greater than the "abandoning criterion". Instead of returning an exact distance, EA returns $+\infty$ if abandoning takes place, which is equivalent to disqualifying the data time series to be the nearest neighbor. By abandoning the full distance computation for unpromising candidates for nearest neighbor, EA significantly accelerates the searching process. 

\textbf{Lower Bounding (LB):} LBs are distance measures that approximate the corresponding elastic measure distance without computing the full distance matrix. The LB distance is always less than or equal to the full elastic measure distance and thus can be used to filter out unpromising candidates in the nearest neighbor search, where full elastic measure distance is not computed for data time series with LB distances greater than existing nearest neighbor. Research efforts have concentrated in developing LBs for DTW, and DTW LBs have inspired the development of several LBs for threshold-based and metric elastic measures. More recently, \cite{paparrizos2023accelerating} proposed the Generalized Lower Bound (GLB) framework that abstracts cost functions of different elastic measures and accumulates the desirable properties of effective LBs, and thereby provides state-of-the-art LBs for all elastic measures.\begin{itemize}
    \item \textbf{LBs for Dynamic Time Warping: } Numerous DTW LBs were proposed in the past two decades. LB\_Kim \cite{kim2001index} is a fast LB with $O(1)$ complexity but rather loose in terms of pruning power. LB\_Keogh  \cite{keogh2004exact} achieves significantly higher pruning power than LB\_Kim by utilizing a warping window and "envelopes": LB\_Keogh first compute the envelope of the query series and compute the distance between the data time series and the query envelope. 
    
    LB\_Keogh inspired several later DTW LBs: LB\_Improved \cite{lemire2009faster} computes LB\_Keogh distance between the data time series against the query time series as well as the query time series against the projection of the data time series; LB\_Enhanced \cite{tan2019elastic} enhances LB\_Keogh by adding alternating bands around the corners of the distance matrix to capture boundary alignment distances;  LB\_Petitjean and LB\_Webb  \cite{webb2021tight} improves the projection computation in LB\_Improved while utilizing bands from LB\_Enhanced to capture boundary distances.
    
\item \textbf{LBs for Threshold-based and Metric Elastic Measures: } Several DTW LBs are adapted to metric and threshold-based elastic measures by adjusting for the different cost functions between these elastic measures and DTW. Modifications were made to LB\_Keogh and LB\_Kim, resulting in the creation of LB\_Keogh-ERP \cite{chen2004marriage} and LB\_Kim-ERP \cite{chen2004marriage}. LB\_Keogh also evolved into LB\_LCSS by substituting DTW’s Euclidean Distance with match and mismatch parameters. 

More recently, \cite{tan2020fastee} introduced lower bounds for TWED and MSM, deriving them based on the unique distance functions of these measures to effectively capture the initial boundary distance and characteristics of the query.

\item \textbf{The Generalized Lower Bounding (GLB) Framework:} the GLB framework \cite{paparrizos2023accelerating} satisfies the key desirable properties of effective LBs: GLB extracts summaries statistics from both the query and the data time series ({\em query and data dependence}), captures the boundary alignments distances ({\em boundary dependence}), and can be reused in future calculations ({\em reusability}).  By abstracting and adapting for the cost functions of different elastic measures, GLB is applicable to all elastic measures and provides state-of-the-art LBs,  including for elastic measures like EDR and SWALE which did not have exisiting LBs in the literature.

    \end{itemize}


%% file: section/S06_sliding.tex
\section{Sliding Measures}
Different from elastic measures, sliding measures evaluate all potential shifts of a time series $X$ along the time dimension to compute the dissimilarity/similarity between every possible shifted $X'$ and another time series $Y$. They define the dissimilarity/similarity between time series $X$ and $Y$ as the minimum/maximum of these calculated values.

\subsection{Shape-based distance (SBD)}
Shape-based distance (SBD) is the distance measure used in k-Shape, the current state-of-the-art algorithm for time series clustering (TSC), introduced by \cite{paparrizos2015k}. Given two univariate time series, $X$ and $Y$ with length $m$, the Coefficient Normalized Cross-Correlation ($NCC_c$) shifts $Y$ relative to $X$ and evaluates the similarity between each shifted result $Y'$ and $X$. Then, the optimal position is determined, denoted as $w$, that maximizes $CC_{w}(X, Y)$ value. To accelerate the computation of $CC$, the $SBD$ employs the Fast Fourier Transform ($FFT$) algorithm and the Inverse Fast Fourier Transform ($IFFT$), as illustrated below:
\begin{equation}
    CC(X, Y) = IFFT(FFT(X) * FFT(Y))
\end{equation}
where $*$ represents taking the complex conjugate in the frequency domain. In the Table \ref{sliding-table}, three types of $NCC$ are identified: biased $NCC$ ($NCC_b$), unbiased $NCC$ ($NCC_u$), and $NCC_c$. In the paper, the authors mentioned that the selection of data normalization methods and Cross-Correlation ($CC$) greatly affects the resulting $CC$ sequence. To determine the most effective $NCC$ variant, the authors conducted distance evaluations using the 1-Nearest Neighbor (1-NN) classifier. They applied three variations of $NCC$ to 48 UCR time-series datasets \cite{dau2019ucr} across three common time-series normalization scenarios and found out that $SBD$, adapting $NCC_c$, significantly outperforms both $NCC_u$ and $NCC_b$. It should be noted that $NCC$ quantifies the similarity between two time series. To derive the distance between them, the $SBD$ is computed using the following equation:
\begin{equation}
    SBD(X, Y) = 1 - max_{w}(\frac{CC_{w}(X, Y)}{\vert\vert{X}\vert\vert\cdot\vert\vert{Y}\vert\vert})
\end{equation}

\subsection{Scaling and translation invariant distance measure (STID)}
K-Spectral Centroid ($K\text{-}SC$) \cite{yang2011patterns} clustering algorithm, modified from k-Means, incorporates a scaling and translation invariant distance measure ($STID$). Given two time series $X$ and $Y$, $STID$ expresses their distance as follows and intends to find the optimal alignment $w$ and the scaling coefficient $\alpha$:
\begin{equation}
    STID(X, Y) = \min_{\alpha, w} \frac{\| X - \alpha Y_w \|}{\| X \|}
\end{equation}
The authors note that the value of $\alpha$ value can be found easily, as the distance function is convex with respect to $\alpha$ for a fixed $w$, as given by the following equation:
\begin{equation}
    \alpha = \frac{X^T Y_w}{\| Y_w \|^2}
\end{equation}
However, there is no simple way to identify the optimal $w$ for $STID$. As suggested by the authors, we can initially determine an alignment $w'$ that synchronizes the peaks of the time series and, then, conduct a localized search for the optimal $w$ within the neighborhood of $w'$.

\begin{table}[h]
\centering
\small
\renewcommand{\arraystretch}{2.5}
\caption{Summary of the sliding measures}
\begin{tabular}[t]{p{0.4\textwidth} p{0.22\textwidth}}
\Xhline{2\arrayrulewidth}
Method &Equations\\
\Xhline{2\arrayrulewidth}
$NCC_b$ \cite{paparrizos2015k}  & $max_{w}(\frac{CC_{w}(X, Y)}{m})$ \\ \hline
$NCC_u$ \cite{paparrizos2015k}  & $max_{w}(\frac{CC_{w}(X, Y)}{m-|w|})$ \\ \hline
$NCC_c$ \cite{paparrizos2015k}  & $max_{w}(\frac{CC_{w}(X, Y)}{\vert\vert{X}\vert\vert\cdot\vert\vert{Y}\vert\vert})$ \\ \hline
$NCC$ \cite{john2020debunking}  & $max_{w}(CC_{w}(X, Y))$ \\ \hline
$STID$ \cite{yang2011patterns} & $\min_{\alpha, w} \frac{\| X - \alpha Y_w \|}{\| X \|}$ \\ 
\Xhline{2\arrayrulewidth}
\end{tabular}
\label{sliding-table}
\end{table}%

%% file: section/S07_kernel.tex
\section{Kernel Measures}
Kernel measures are functions that compute the distance between two time series by implicitly mapping those data points to another, potentially higher-dimensional, space using a kernel operation.  Mathematically, a kernel is defined as:
\begin{align*}
    k &: \mathcal{X} \times \mathcal{X} \rightarrow \mathbb{R} \\
    k(X, Y) &= <\Phi(X), \Phi(Y)>,
\end{align*}
where $X, Y \in \mathcal{X}$ denotes two time series and  $<,>$ represents an inner product. $\Phi$ represents the mapping to a high dimensional Hilbert space \cite{vapnik,scholkopf2000kernel}.

To guarantee the existence of such a Hilbert Space with desirable geometric properties, e.g., orthogonality, projection \cite{ong}, it is suggested to employ positive definite kernels under Mercer's theorem \cite{mercer1909functions}: A symmetric kernel function \(k: \mathcal{X} \times \mathcal{X} \rightarrow \mathbb{R}\) is said to be positive definite (PD) if for all $n\in \mathbb{N}$, the $i$th instance \(X^i \in \mathcal{X}\), the following condition holds for all real number $c_i$, $c_j$ ($1\leq i, j \leq n$):
\begin{align*}
    \sum_{i=1}^{n}\sum_{j=1}^{n} c_i c_j k(X^i, X^j) \geq 0.
\end{align*}


\subsection{Gaussian Kernels}
The popular Radial Basis Function (RBF) \cite{shawe-taylor} is an instance of the general gaussian kernels that exploits the Euclidean Distance. RBF is formally defined in Table \ref{tab:kernel}, where $x, y$ are input time series, and $\sigma$ is a hyper-parameter. In prior works  \cite{haasdonk2004learning}, RBF has been proven to be positive definite and widely used in diverse downstream tasks \cite{lei2007study}. Prior studies also explore the combination of Gaussian kernel with other measures such as elastic measures \cite{bahlmann2002online,shimodaira2001dynamic}. However, in previous studies \cite{lei2007study, gudmundsson2008support}, it is shown that Gaussian elastic measures such as Gaussian DTW (GDTW) kernel do not guarantee positive definite symmetric and GDTW cannot outperform RBF in the support vector machine (SVM) framework. It suggests that further analysis and extensive extension work are needed to explore this direction \cite{zhang2010time}.

\subsection{(Log) Global Alignment Kernel}
Global Alignment Kernel (GAK) \cite{cuturi} considers all alignments between sub-sequences of two series rather than focusing only on the optimal alignment path. Under favorable conditions, GAK has been proven to be positive definite. Since the diagonal elements in GAK show dominance when compared to other off-diagonal elements, Log Global Alignment Kernel (LGAK) uses logarithmic operation to alleviate this issue, showing superior performance in classification tasks \cite{cuturi2011fast}. LGAK is defined in Table \ref{tab:kernel}, where $X, Y$ are input time series with a length of $n$ and $m$. $\pi$ represents an alignment operation between two sequences, and $\sigma$ is a hyper parameter. $\mathcal{A}(n,m)$ denotes the set of all possible alignments.

\subsection{Kernel Dynamic Time Warping}

In prior studies, it is shown that many elastic distances do not have definite kernels as an extension in the kernel family \cite{gibet}. It is noted that the possible cause of indefiniteness in elastic measure kernels is the presence of $min$ and $max$ operations in their recursions. To solve this issue, Kernel Dynamic Time Warping (KDTW) \cite{gibet} is proposed as a dissimilarity measure constructed from DTW with a positive definite kernel, which replaces the $min$ and $max$ operator with a summation ($\sum$) operator and iterate through all alignment paths.
The formula for KDTW is shown in Table \ref{tab:kernel},
where $\Delta_{i,j}$ is the Kronecker's symbol and $h$ is a symmetric binary non-negative function (usually between 0 and 1)
$\sigma \in \mathbb{R^+}$ denotes a bandwidth parameter which weights
and $\beta \in [\frac{1}{3}, 1]$ is for normalization.
In addition, KDTW can be generalized to Recursive Edit Distance Kernel (REDK), which could benefit from PD kernels with other elastic measures including ERP and TWED. 

\subsection{Shift Invariant Kernel}
Shift Invariant Kernel (SINK) \cite{paparrizos2019grail} computes the distance between time series $X$ and $Y$ by summing all weighted elements of the Normalized Cross-Correlation ($NCC$, see Section 4) sequence between $X$ and $Y$. Formally, SINK is defined in Table \ref{tab:kernel},
where $\gamma > 0$ is the bandwidth parameter that determines weights for each inner product $<X, Y>$. Considering the scaling problem of the off-diagonal values, a normalization strategy is performed to alleviate the issue.

\begin{table}[t]
\caption{Summary of the kernel measures.}\label{tab:kernel}
\resizebox{0.8\columnwidth}{!}{
\begin{tabular}{p{0.15\textwidth}p{0.5\textwidth}}
\Xhline{2\arrayrulewidth}
Method & \makecell[c]{\qquad\qquad\qquad\qquad\qquad Formula} \\

\Xhline{2\arrayrulewidth}

RBF \cite{shawe-taylor}      & {
\begin{align*} k(X, Y) = \exp(-\frac{\norm{X-Y}^2}{2\sigma^2}) \end{align*}} \\ 
\hline
\addlinespace[0.2cm]
LGAK \cite{cuturi}     & {
\begin{align*}LGAK(X, Y,\sigma)=\sum_{\pi\in\mathcal{A}(n,m)} \prod_{i=1}^{|\pi|} \exp\left(-\frac{\norm{x_{\pi_1(i)} - y_{\pi_2(i)}}^2}{2\sigma^2}-\log\left(2-\exp\left(-\frac{\norm{x_{\pi_1(i)} - y_{\pi_2(i)}}^2}{2\sigma^2}\right)\right)\right)\end{align*}}    
\\ 
\addlinespace[0.2cm]
\hline
\addlinespace[-2cm]
\makecell[l]{ \\ \\ \\ \\ \\ \\ \\ \\ \\ \\ \\KDTW \cite{gibet}} & {
\begin{align*}
    k(X_i,Y_j,\sigma) &= e^{- (X_i-Y_j)^2/\sigma } \\
    KDTW^{xy}(X_i,Y_j,\sigma) &= \beta \cdot k(X_i,Y_j,\sigma) \cdot \sum
        \begin{cases}
            h(i-1,j)KDTW^{xy}(X_{i-1},Y_j) \\ h(i-1,j-1)KDTW^{xy}(X_{i-1},Y_{j-1}) \\ h(i,j-1)KDTW^{xy}(X_i,Y_{j-1}) \\
        \end{cases}\\
        KDTW^{xx}(X_i,Y_j,\sigma) &= \beta \cdot \sum 
        \begin{cases}
        h(i-1,j) KDTW^{xx}(X_{i-1},Y_j) \cdot k(X_{i},Y_i,\sigma) \\ \Delta_{i,j} h(i,j)KDTW^{xx}(X_{i-1},Y_{j-1}) \cdot k(X_i,Y_j,\sigma) \\ h(i,j-1)KDTW^{xx}(X_i,Y_{j-1}) \cdot k(X_j,Y_j,\sigma) \\
        \end{cases} \\
        KDTW(X,Y) &= KDTW^{xy}(X,Y) + KDTW^{xx}(X,Y) \\ 
\end{align*}
} \\
\addlinespace[-0.5cm]
\hline
\addlinespace[-2cm]
\makecell[l]{ \\ \\ \\ \\ \\ \\ \\SINK \cite{paparrizos2019grail}} & {\begin{align*}
     & \\
    k_s(X,Y,\gamma) &= \sum e^{\gamma  NCC(X,Y)} \\
    SINK(X,Y,\gamma) &= \frac{k_s(X,Y,\gamma)}{\sqrt{k_s(X,X,\gamma) \cdot k_s(Y,Y,\gamma)}}
\end{align*}} \\

\Xhline{2\arrayrulewidth}
\end{tabular}}
\end{table}


%% file: section/S08_feature.tex
\section{Feature-based Methods}
Feature-based methods are proposed to identify descriptive attributes that globally represent the characteristics of time series and perform distance computation based on the extracted features. Given the difficulty in separating feature-based distance measures from their downstream tasks and the multitude of such tasks, this survey will mainly concentrate on exploring how feature-based distance measures are employed in TSC, a prevalent downstream task, and involves grouping similar time series data according to specific criteria without any supervision. In this survey, we focus on discussing key feature-based methods in detail, given the multitude of features and their combinations that are related to specific tasks. Related methods can be found in Table \ref{Feature-distance_table}.

Driven by the observation that long time series and missing data often lead to failures in numerous existing clustering algorithms, Characteristic-Based Clustering (CBC) \cite{wang2006characteristic} was proposed. CBC employs measures of global structural characteristics, combining classical and advanced statistical features, to cluster time series. These characteristics include trend, seasonality, periodicity, serial correlation, skewness, kurtosis, chaos, non-linearity, and self-similarity. Leveraging these global representations, CBC effectively reduces the dimensionality of time series data and enhances its robustness against missing or noisy data. The authors emphasized that selecting an appropriate set of features can enhance computational efficiency and improve clustering outcomes, and they developed a new method based on a greedy Forward Search (FS) algorithm to identify the optimal subset of features. In the paper, CBC does not mandate specific distance measures, allowing for integration with various distance measures as needed, and the selection of a distance measure is customized to meet the specific needs of the task being addressed.

Recognizing the time-consuming nature of feature engineering in time series analysis, which involves the challenging task of sifting through a vast array of signal processing and time series analysis algorithms to identify relevant and significant features, and tending to accelerate this process, tsfresh \cite{christ2018time}, a widely used and well-known python package, was purposed. Tsfresh provides 63 methods for time series characterization, yielding 794 distinct time series features by default, and also automates the feature extraction and selection through the FeatuRe Extraction based on Scalable Hypothesis tests (FRESH) algorithm \cite{christ2016distributed}. This algorithm demonstrates scalable performance, linearly increasing with the number of features, the quantity of devices/samples, and the number of different time series involved. However, because the computational costs of features vary based on their complexities, modifying the set of features computed by tsfresh can significantly impact its total execution time. The tsfresh paper does not specify a preferred distance measure for evaluating the distances between extracted features. Numerous combinations of features and distance measures can be explored, depending on the specific needs and the nature of the data.

\begin{table}[!htp]
\footnotesize
\caption{Summary of the Feature-based Measures}
\resizebox{0.55\linewidth}{!}{
\begin{tabular}[t]{lccc}
\Xhline{2\arrayrulewidth}  \addlinespace[0.2cm]
Method &Feature &Distance &Dim\\

\Xhline{2\arrayrulewidth} \addlinespace[0.2cm]
TSS-IOF-ED \cite{alcock1999time} &First-Order, Second-Order & ED &I \\
TSC-GC-ED \cite{wang2005dimension} &Global &ED &I \\ 
CBC \cite{wang2006characteristic}  & Comprehensive & * & I \\
TSC-SSF \cite{wang2007structure}  & Statistical & * & M \\
TSBF \cite{baydogan2013bag} & Statistical & ED &M \\
FEDD \cite{cavalcante2016fedd} & Statistical & Cosine, Pearson &I \\
FBC \cite{afanasieva2017time}  & Fuzzy & ED &I \\ 
hctsa \cite{fulcher2017hctsa}  & Comprehensive & * & I\\
tsfresh \cite{christ2018time}  & Comprehensive & * &M \\
catch22 \cite{lubba2019catch22}  & Canonical & * &M \\
TSC-CN \cite{bonacina2020time}  & Visibility Graph & ED &M \\
FeatTS \cite{tiano2021featts,tiano2021feature} & TSfresh & ED & I \\
TSC-GPF-ED \cite{hu2021classification} & Global, Peak & ED &I \\
TSC-FDDO \cite{zhang2021fault} & Comprehensive & ED &I \\
Time2Feat \cite{bonifati2022time2feat} &Comprehensive &* &M \\
theft \cite{henderson2022feature} & Comprehensive &ED &I \\
AngClust \cite{li2022angclust}  & Angular & PC &M \\
FGHC-SOME \cite{wunsch2022feature}  & Statistical & PC &M \\
TSC-VF \cite{wu2023imaging}  & Visual & * & I \\
FTSCP \cite{enes2023pipeline} & Comprehensive & ED &M \\

\addlinespace[0.2cm]
\Xhline{2\arrayrulewidth} \addlinespace[0.2cm]
\end{tabular}}
    \begin{tablenotes}
      \scriptsize
      \centering
      \item  I: Univariate; M: Multivariate; *: Arbitrary Distance;
     \end{tablenotes}
\label{Feature-distance_table}
\end{table}%

The 22 CAnonical Time-series CHaracteristics (catch22) \cite{lubba2019catch22} is a set of features derived from the highly comparative time-series analysis (hctsa) \cite{fulcher2017hctsa} toolbox. Originally hctsa provided a vast collection of over 7,700 time-series features, with a reduced, filtered set of 4,791 features. The catch22 set further distills this extensive list to 22 essential features, offering a more manageable and focused approach for analyzing time-series data. The hctsa tool is capable of selecting suitable features for specific applications, but this process is computationally intensive and involves unnecessary redundant calculations. Based on their findings, the developers of catch22 established a systematic pipeline that incorporates statistical prefiltering to identify important features, performance filtering to evaluate the computational efficiency and discriminative effectiveness of the features, and redundancy minimization to remove repetitive information. It is worth mentioning that the authors employ Pearson Correlation (PC) as the distance measure for hierarchical clustering with complete linkage during redundancy minimization, selecting one representative feature from each cluster to form a canonical feature set. Implementing this pipeline yielded a standardized set of 22 features, significantly enhancing computational efficiency and scalability with a minimal average loss of just 7\% in classification accuracy. Furthermore, catch22 effectively encapsulates the varied and representative attributes of time series. Its refined features are classified into 7 categories: distribution, simple temporal statistics, linear and nonlinear autocorrelation, successive difference, fluctuation analysis, and others. After feature extraction with catch22, the choice of distance measure is flexible and should be tailored to the specifics of the data and the analytical task.

%% file: section/S09_model.tex
\section{Model-based Methods}


One of the primary features of model-based distances is the ability to integrate prior knowledge about the data-generating process into the assessment of similarity/ dissimilarity. Model-based approaches concentrate on representing the underlying distribution of time-series data through a collection of parameters. Thus, the distance between two time series can be expressed as a comparison of the parameter sets associated with each. Representative modeling techniques in time series analysis are Gaussian Mixture Model (GMM), Hidden Markov Model (HMM), AutoRegressive Moving Average (ARMA), and AutoRegressive Integrated Moving Average (ARIMA). For the same reasons outlined in the feature-based methods section, this survey primarily concentrates on gathering instances of how model-based methods are applied in TSC and related methods can be found in Table \ref{model-distance_table}.\\

\noindent\textbf{Gaussian Mixture Model (GMM)} \\
The GMM \cite{bishop2006pattern, green2019introduction} is a probabilistic model that represents a univariate time series $X$ with length $n$ as a combination of multiple Gaussian distributions. Each Gaussian distribution $j$ in the GMM is denoted as $\mathcal{N}(x_i| \mu_j, \sigma_{j}^2)$ with mean $\mu_j$ and variance $\sigma_{j}^2$. Associating the mixture proportion $\pi_j$ with $j^{th}$ Gaussian component, the function of the GMM mixed by $K$ Gaussian distribution is obtained by the following equations:
\begin{align}
     \mathcal{N}(x_i | \mu_j, \sigma_j^2) &= \frac{1}{\sqrt{2\pi\sigma_j^2}} \exp\left(-\frac{(x_i - \mu_j)^2}{2\sigma_j^2}\right) \\
     p(X) &= \prod_{i=1}^{n} \sum_{j=1}^{K} \pi_j \mathcal{N}(x_i | \mu_j, \sigma_j^2)
\end{align}

The parameters of a GMM are learned through the Expectation-Maximization (EM) algorithm. The application of a model-based distance measure, post the GMM learning, is exemplified in the work referenced in \cite{tran2002fuzzy}. Observing and analyzing the false acceptance error caused by existing normalization methods for speaker's score computation in the speaker verification field, this paper's normalization approach is based on fuzzy c-means (FCM) clustering. This new method challenges the existing assumption that all background speakers contribute equally in terms of their likelihood values. The FCM algorithm employs fuzzy partitioning to determine a membership matrix, which calculates the degree of membership of each time series belonging to a specific cluster. To optimize the fuzzy objective function, the FCM membership function is expressed as a function of distance. Modified from the FCM membership function, the authors introduce their formula for calculating the FCM membership score. They redefine the concept of distance within the fuzzy membership function as $d^2 (X, \lambda_i) = -\log P(X | \lambda_i)$, where $ P(X | \lambda_i)$ denotes the GMM likelihood function and indicates the likelihood of a time series $X$ being generated by the trained GMM model labeled as $\lambda_i$. After obtaining the FCM membership score, the next step involves assessing the FCM membership score against a predefined decision threshold to determine the outcome.\\

\noindent\textbf{Hidden Markov Model (HMM)} \\

HMM \cite{bishop2006pattern}, supported by two primary assumptions, is a type of probabilistic graphical model designed to deduce the latent hidden states and the transitions among these states using the observed data. Given the current state $S_i$, the first assumption mentions that the next state $S_{i+1}$ is solely influenced by $S_i$, whereas the second assumption states that the current observation $O_i$ is exclusively dependent on $S_i$. In HMM, three key probabilities are defined and need to be learned during the training stage: initial probability, transition probability, and emission probability. The initial probability is a set of probabilities indicating how likely it is for the sequence to start in each possible hidden state. Transition probability represents the likelihood of moving from one hidden state to another within the model. Finally, emission probability specifies the likelihood that a specific observation is generated by a given hidden state. HMM learning also utilizes the EM algorithm to estimate parameters, with one of the most renowned implementations being the Baum-Welch algorithm \cite{baum1972inequality}.

An example of applying model-based distance to assess the distance is illustrated in the study \cite{ghassempour2014clustering}. The authors propose a method where, instead of directly computing the distance between two time series, each time series is transformed into an HMM model. They then introduce and employ a distance measure called symmetrized KL divergence (S-KL Divergence) to determine the distance between the two resulting models. The S-KL Divergence proposed by authors balances the simplicity of one-point approximation and the comprehensiveness of full Monte Carlo approximation of KL divergence. Assuming the observed dataset sufficiently captures the variety of possible time series, the authors feed the approximated HMM probability densities of two distinct models into their proposed distance measure to evaluate the distance between these models. Utilizing the derived distance matrix, the authors conduct HMM clustering which ultimately leads to the clustering of the time series, a downstream task of distance measures. This process naturally groups together time series with similar characteristics, operating without the need for supervision.\\

\noindent\textbf{Autoregressive Model (AR) \& Moving Average Model (MA)}\\

The AR model, referenced in \cite{hyndman2018forecasting}, is grounded in the concept that the value $x_t$ in a univariate time series at time $t$ is derived from a linear combination of its $p$ preceding data points. Considering $\phi_i$ as the coefficient associated with $x_{t-i}$, $\varepsilon_t$ as the error term at time $t$, and $c$ as the constant term, $x_t$ is computed using the equation:
\begin{equation}
    x_t = c + \phi_1 x_{t-1} + \phi_2 x_{t-2} + \ldots + \phi_p x_{t-p} + \varepsilon_t
\end{equation}

Establishing a connection between the value of $x_t$ and past $q$ errors, the MA model \cite{hyndman2018forecasting} is another classic statistical model. Denoting $\theta_i$ as the parameter associated with $\varepsilon_{t-i}$ and $\varepsilon_t$ as the error at time $t$, $x_t$ can be computed as the sum of the time series mean $\mu$ with a linear combination of previous errors and follows the exact equation as below:
\begin{equation}
    x_t = \mu + \theta_1 \varepsilon_{t-1} + \theta_2 \varepsilon_{t-2} + \ldots + \theta_q \varepsilon_{t-q} + \varepsilon_t
\end{equation}

Integrating the AR with MA yields the Autoregressive Moving Average (ARMA) model \cite{hyndman2018forecasting}. This model calculates the value $x_t$ by considering both its past values and preceding error terms. Building upon the ARMA framework, the Autoregressive Integrated Moving Average (ARIMA) \cite{hyndman2018forecasting}, incorporates differencing into the ARMA structure, thereby enhancing its ability to analyze non-stationary time series data whose statistical properties change over time.

\cite{piccolo1990distance} proposes a parametric method for the class of ARIMA invertible models, utilizing the ED calculated from their expanded autoregressive forms. After learning the ARIMA models for two time series $X$ and $Y$, the coefficients for each are retained in the sequences $\pi_x$ and $\pi_y$ respectively and, then, the distance between these two ARIMA models can be quantified by calculating the ED between $\pi_x$ and $\pi_y$. Additionally, beyond employing the ED to determine the distance between two fitted AR expansions, one may also apply a hypothesis test \cite{maharaj2000cluster} and log-likelihood \cite{xiong2002mixtures} to assess it. 

\begin{table}[!htp]
\footnotesize
\caption{Summary of the model-based distance measures.}
\label{model-distance_table}
\resizebox{0.6\linewidth}{!}{
\begin{tabular}[t]{lccc}
\Xhline{2\arrayrulewidth}  \addlinespace[0.2cm]
Method &Model &Distance Measure &Dim\\

\Xhline{2\arrayrulewidth} \addlinespace[0.2cm]
TSC-ARIMA-ED \cite{piccolo1990distance} &ARIMA &ED &I \\
TSC-D-HMM \cite{li1999temporal} &HMM &Log-likelihood &M \\
ICL \cite{biernacki2000assessing} &GMM &Log-likelihood &M \\
TSC-AR-HT \cite{maharaj2000cluster} &AR & Hypothesis test &I \\
MBCD \cite{ramoni2000multivariate} &Markov Chain &KL distance &M \\
TSC-LPC-ARIMA \cite{kalpakis2001distance} &ARIMA &ED &I \\
BHMMC \cite{li2001building}  &HMM &BIC &M \\
FCM-SV \cite{tran2002fuzzy} &GMM &Log-likelihood &I \\
HMM-TWM \cite{wang2002hidden}  & HMM &ED &I \\
TSC-ARMAM \cite{xiong2002mixtures} &ARMAs &Log-likelihood &I \\
CLUSTSEG \cite{same2011model} & Regression Mixture & $L_2$ distance &I \\
LMAR, LMMAR \cite{kini2013large} &LMAR, LMMAR &Mahalanobis &I\\
TSC-HMM-S-KL \cite{ghassempour2014clustering} &HMM &S-KL divergence &M \\
MV-ARF \cite{tuncel2018autoregressive} & AR Ensembles & MSE & M \\
K-MODELS \cite{hoare2022k} & ARMA, ARIMA & K-Models loss &I \\
\Xhline{2\arrayrulewidth} \addlinespace[0.2cm]
\end{tabular}}
    \begin{tablenotes}
        \centering
      \scriptsize
      \item  I: Univariate; M: Multivariate;
     \end{tablenotes}
\end{table}%

%% file: section/S10_embedding.tex
\section{Embedding-based Methods}
\label{sec:emb}

As defined in Sec. \ref{sec:taxo}, Embedding measures focus on constructing new representations of time series, and then capturing the dissimilarity information from this embedded representation, e.g., using a simple lock-step distance measure like ED. Although the pipeline looks similar to other distance measures such as feature-based and model-based, embedding measures show intrinsic differences when compared with the other two: 1) Extensive exploration and calculation of statistical features are not required for embedding measures. In other words, embedding measures capture the characteristics information in a latent space. 2) compared with model-based measure, embedding measures do not need an explicit model to learn the distribution, where the representation ability would be limited by the model itself.
In the following, we are going to review four widely-used embedding measures: Generic RepresentAtIon Learning (GRAIL) framework \cite{paparrizos2019grail}, Random Warping Series (RWS) \cite{wu2018random}, Shift-invariant Dictionary Learning (SIDL) \cite{zheng2016sidl}, Similarity Preserving Representation Learning method (SPIRAL) \cite{lei2019similarity}. We will also introduce recent work discussing the use of deep neural networks as tools for embedding time series. We include related embedding-based methods in Table \ref{embedding-distance_table}.

\subsection{GRAIL: Generic RepresentAtion Learning }
The Generic RepresentAtIon Learning (GRAIL) framework \cite{paparrizos2019grail} builds compact representations of time series which preserve the properties of a user-specified comparison function. GRAIL combines the Shift-invariant Kernel (SINK) with the Nystrom method for low-rank approximation of the Gram matrix. GRAIL selects landmark time series, which are chosen through a clustering approach, to perform this approximation of the data using the distance between the data and the landmark series. GRAIL makes use of many approximations and optimization steps to help ensure the computational feasibility of the representation learning process and provides an unsupervised solution for tuning important and needed kernel parameters without the need for computationally infeasible supervised tuning. 
\subsection{RWS: Random Warping Series}
Random Warping Series \cite{wu2018random} propose learning representations of time series via computing the alignment of the series in a given dataset and a random distribution of time series. The work shows that this alignment can be approximated by computing the alignment of a finite number of sampled time series drawn from the distribution, making the computation requirements relatively much lower than is seen in explicit kernel computations such as GAK. The work also shows that positive definite kernels can be defined over this space for learning with SVMs. The authors also provide proof that the RWS representation method preserves the GAK.
\subsection{SPIRAL: Similarity Preserving Representation Learning}
Similarity Preserving Representation Learning \cite{lei2019similarity} is a representation method which also relies on reducing the complexity of full similarity matrix computation using random sampling. First, a similarity measure based on DTW is introduced to measure the similarity between two series in a numerically stable way. SPIRAL avoids a full distance matrix computation for DTW, which is a prohibitively expensive operation in many real applications. This is done by sampling $\mathcal{O}(\textit{n}log(\textit{n}))$ pairs of time series from the dataset and observing their pairwise similarity. This gives a partially observed distance matrix for DTW on the data. A representation is then constructed for all $\textit{n}$ time-series by using this partially observed matrix to learn a new optimal feature representation whose inner product approximates the pairwise DTW similarities. The proposed method for learning this feature representation is via a non-convex optimization process for factorization of the partially observed matrix. This may incur a high computational cost if solved naively, though the authors provide a fast and parameter-free approach to computing this factorization.

\begin{wraptable}{r}{0.6\textwidth}
\vspace{-0.5cm}
\footnotesize
\caption{Summary of the embedding-based methods.}
\label{embedding-distance_table}
\resizebox{\linewidth}{!}{
\begin{tabular}[t]{lll} 
\Xhline{2\arrayrulewidth} 
Method & Embedding Model & Dim\\

\Xhline{2\arrayrulewidth} \addlinespace[0.2cm]
GRAIL \cite{paparrizos2019grail} & SINK kernel similarity \& Spectral decomposition & I \\
RWS \cite{wu2018random} & Randomized DTW Similarity & I \\
SPRIAL \cite{lei2019similarity} & DTW distance matrix approximation & I \\
SIDL \cite{zheng2016sidl} & Dictionary learning \& Sparse coding & I \\
Autoencoder \cite{bank2021autoencoders} & Generic DNN backbone & M \\
Time2Vec \cite{kazemi2019time2vec} & Single learnable layer \& periodic activation & I \\
TS2Vec \cite{yue2022ts2vec} & Dilated CNN and hierarchical  & M \\
LLM Encoding \cite{jin2024position} & Multimodal LLM & M \\

\Xhline{2\arrayrulewidth} \addlinespace[0.2cm]
\end{tabular}}
    \begin{tablenotes}
      \scriptsize
      \centering
      \item  I: Univariate; M: Multivariate;
     \end{tablenotes}

\vspace{-0.5cm}
\end{wraptable}

\subsection{SIDL: Shift-invariant Dictionary Learning}
Shift-invariant Dictionary Learning \cite{zheng2016sidl} approaches the representation problem by learning a shift-invariant basis or dictionary over the data using an unsupervised approach. This is done by optimizing a dictionary of relevant shift invariant patterns in the dataset, which may be of length less than the length of the series, jointly with a sparse coding of the data. This joint optimization efficiently learns a high-quality representation of the input data by alternating between optimization of the sparse coding, and optimization of the dictionary. In doing so the problem is reduced from being non-convex to two smaller problems that are well constrained and can be efficiently solved.

\subsection{Deep Learning}
Deep Neural Networks (DNNs) excel at extracting high-level and hierarchical transformations from many data types. This is done by repeated weighted application of parameterized hidden units equipped with non-linear activation functions to create transformations to novel spaces where the data may be more effectively characterized. This strong capability to learn abstract representations has led to work suggesting the latent space of deep-learning architectures can be used as an encoder to generate representation methods for time series data. A diverse range of methods have been proposed for this task, with diverse architectural considerations such as bespoke clustering loss terms or triplet losses. In the clustering context, these have been reviewed comprehensively in a recent work \cite{lafabregue2022dlrepresentation}. The authors find that while there is high variability in clustering results due to architecture selections, embeddings from DNNs employed as encoders do achieve performance improvements in embedding time series for the clustering task. This suggests further that Deep Learning models can construct encoders to find meaningful time series embeddings for other important tasks provided proper loss functions and training steps. These embeddings from the clustering context are naturally applicable to distance-based classification with the Euclidean distance. The work finds that Convolutional Neural Networks (CNNs) \cite{imagenet2012} \cite{lecun_98} perform the best for extracting features but that in principle any deep learning backbone can be used, leaving broad flexibility for adapting new architectures and emerging deep learning components.

This area of research is rapidly evolving and we shall highlight some exemplary methods and briefly introduce their architectures. Architectures widely used in other areas can be adapted to time series and are commonly used in the context of an encoder or autoencoder \cite{bank2021autoencoders}. Long-Short Term Memory Networks (LSTMs) \cite{bengioLSTM} are a classic form of recurrent network to handle sequence modeling tasks and solve the vanishing gradient problem in modeling long sequences. A recent work \cite{Succetti_2023} addresses some of the weaknesses of LSTMs using a new architecture making use of bi-directional LSTMs.  CNNs, first popularized in computer vision and image processing tasks can be adapted to 1-Dimensional filters \cite{Ismail_Fawaz_2019} to learn features in a non-recurrent context over a time series. Time2vec \cite{kazemi2019time2vec} provides a simplified learnable method for generating embeddings of time features using periodic activation functions. A convolutional architecture that is scalable with respect to the lenght of the time series is proposed in \cite{franceschi2019unsupervised}, by a stacked convolutions which are made sparse by dilating them exponentially wide across time at each layer. TS2vec \cite{yue2022ts2vec} is based on another dilated CNN architecture and proposes novel hierarchical contrasting and textual consistency learning components to generate time series embeddings which achieve superior performance in unsupervised tasks of anomaly detection and forecasting.

Apart from learning an embedding from a single modality of time series, many researchers also delve into the exploration of multi-modal learning. This approach integrates the time series analysis with additional data sources like images, text, and event-related information \cite{CHENG2022108218,ekambaram2020attention,sun2023test,jin2024position}. The goal is to enhance the model learning process by leveraging diverse information from modalities other than the time series itself, which leads to more robust models. In recent years, the success of large language models (LLM) such as GPT-4 and LLaMA \cite{touvron2023llama} have attracted great attention to bring the human prior knowledge from text-based data sources to time-series domain \cite{sun2023test,chang2023llm4ts,garza2023timegpt1,jin2024position,ansari2024chronos}. By aligning the embedding space of time series with natural language, new models are enabled to learn a robust representation that leverages the strengths of LLMs while capturing the intrinsics of time-series modality \cite{jin2024position}.

%% file: section/S11_mtsdist.tex
\section{Multivariate Time-series Distance Measure}
\label{sec:multivariate}
As defined earlier in Section~\ref{sec:preliminaries}, multivariate time series (MTS) are collections of univariate time series that measure different signals at the same resolution, typically sourcing from the same physical object or process~\cite{yang04}.
These different signals are referred to as the \textit{dimensions}, \textit{channels}, or \textit{variates} of the MTS.
While these terms can be used interchangeably, we use the notion of channels in this survey for consistency.
Examples of MTS include motion capture data (e.g., the position and acceleration of different body parts)~\cite{alon03,corradini01,kadous02}, data from climate sensors (e.g., a sensor array measuring the temperature, humidity, and air pressure, at a certain location)~\cite{liess2017}, and Electro Encephalogram (EEG) data in medicine~\cite{zhang95}.

\subsection{Extending Univariate Distances to the Multivariate Case}\label{sec:multivariate_intro}
In principle, distance measures for MTS are no different from those for univariate time series; they also aim to quantify the dissimilarity between two MTS.
Moreover, many existing distance measures are natural extensions of the distance measures discussed earlier throughout this chapter. 
For example, the Euclidean distance for MTS essentially involves computing the squared differences between all corresponding points \emph{and channels} of two MTS, summing them up, and taking the square root of the result.
Extension of all other lock-step measures is similarly straightforward.
However, the additional dimensionality in the form of channels introduces new challenges and opportunities when extending other types of distance measures, such as elastic measures, sliding measures, and embedding measures.
Particularly, when looking to extend these measures to the multivariate case, the question arises if the measure-specific alignments, operations, transformations, and models should be applied to each channel separately, or if they should be applied to the MTS as a whole.
For instance, in the case of DTW, it is non-trivial to decide whether the alignment (i.e., warping path) should be shared among all channels or if one allows each channel pair to have its own alignment.

Yekta et al.~\cite{yekta17} coined these two strategies of time series alignment for DTW as \emph{dependent} (i.e., one shared alignment) and \emph{independent} (i.e., channel-specific alignments) alignment.
Interestingly, the authors demonstrated through real-world examples that the choice of alignment type is data and context-dependent, and can have a significant impact on the results.
As an example, they consider the task of clustering RGB images (i.e., MTS with 3 channels with the pixels as the time dimension), and describe two scenarios; (a) a case where two similar paintings are drawn at different times, resulting in different hues of the same color, and (b) a case where two identical photos are taken with different lighting conditions, resulting in the same hues of different colors.
The MTS are similar in both cases. However, the first case will involve the colors having varying offsets (i.e., requiring independent alignment), while the second case will involve a fixed offset between the colors (i.e., requiring dependent alignment).
As such, they conclude that the optimal alignment type depends on the context of the data.

Here, we argue that this distinction between channel-dependent and channel-independent alignment in MTS distance measures is not only relevant for DTW, or elastic measures in general, but can be applied to all categories of multivariate measures.
Particularly, we extend the concept of alignment types to pre-processing and modeling of MTS in general, referring to it as a distance measure's \emph{dependency model}.

\textbf{Dependency Models:}
The dependency model of an MTS distance measure captures the assumed dependencies between the channels of the MTS, and therefore impacts how measure-specific alignments, operations, transformations, and models are performed or fitted over the channels. This can be done in either (a) a channel-dependent manner, meaning the MTS is treated in a holistic manner and the operations are performed over all channels of the MTS simultaneously, or (b) in a channel-independent manner, meaning the channels of the MTS are treated in an atomic manner and the operations are performed over each channel separately.

Univariate distance measures can be extended to multivariate measures in both a channel-dependent and a channel-independent fashion, with the exact definition of each variant being specific to the measure's type (i.e., elastic, sliding, embedding, etc.).
Measures that do not involve any time alignment, transformation, or model fitting, such as lock-step measures are naturally channel-independent and therefore only have one multivariate variant.
In the following, we will discuss the definition of both dependency models for each measure type.

In general, if a measure is channel-dependent it implies that a single alignment, transformation, or model is derived for the whole MTS. 
Channel-independent variants of measures generally treat each channel as a univariate time series and perform univariate operations on each channel separately, leading to a separate distance for each pair of corresponding channels between two MTS. A total distance is then computed by summing these pairwise distances. Alternatively, while not covered in existing literature, it is possible to compute a final distance through other aggregation methods, such as taking the maximum or the average of the pairwise distances.

\subsection{Multivariate Lock-Step Measures}
As lock-step measures treat each time point independently, their extensions to multivariate time series naturally involve \emph{concatenation of the channels} into a single univariate time series, and then applying the univariate measure to the resulting time series. This implies that all multivariate lock-step measures are inherently channel-independent.
For Euclidean distance, this means that the distance between two MTS $\vec{X}$ and $\vec{Y}$ is computed as:
\begin{equation}
    L_2(\vec{X}, \vec{Y}) = \sqrt{\sum_{j=1}^{c} \sum_{i=1}^{n} (\vec{X}_i^{(j)} - \vec{Y}_i^{(j)})^2}
\end{equation}
The extension of other lock-step measures follows from the same principle. 
For example, the extension of the general Minkowski distance is defined as:
\begin{equation}
    L_p(\vec{X}, \vec{Y}) = \left(\sum_{j=1}^{c} \sum_{i=1}^{n} |\vec{X}_i^{(j)} - \vec{Y}_i^{(j)}|^p\right)^{\frac{1}{p}}
\end{equation}
where $p \geq 1$ is the order of the Minkowski distance.
This also holds for more specialized lock-step measures such as the inner product and the cosine similarity, which are defined as:
\begin{equation}
    \langle \vec{X}, \vec{Y} \rangle = \sum_{j=1}^{c} \sum_{i=1}^{n} \vec{X}_i^{(j)} \cdot \vec{Y}_i^{(j)}
\end{equation}
\begin{equation}
    \cos(\vec{X}, \vec{Y}) = \frac{\sum_{j=1}^{c} \sum_{i=1}^{n} \vec{X}_i^{(j)} \cdot \vec{Y}_i^{(j)}}{\sqrt{\sum_{j=1}^{c} \sum_{i=1}^{n} (\vec{X}_i^{(j)})^2} \cdot \sqrt{\sum_{j=1}^{c} \sum_{i=1}^{n} (\vec{Y}_i^{(j)})^2}}
\end{equation}
As can be seen, the extension of lock-step measures flows naturally from the concatenation of the channels into a single univariate time series.
For this reason, we will not provide explicit definitions for the multivariate extensions of all lock-step measures discussed in Section~\ref{sec:lock-step}.

\subsection{Multivariate Elastic Measures}
\begin{figure}[t]
    \centering
    \includegraphics[width=\linewidth]{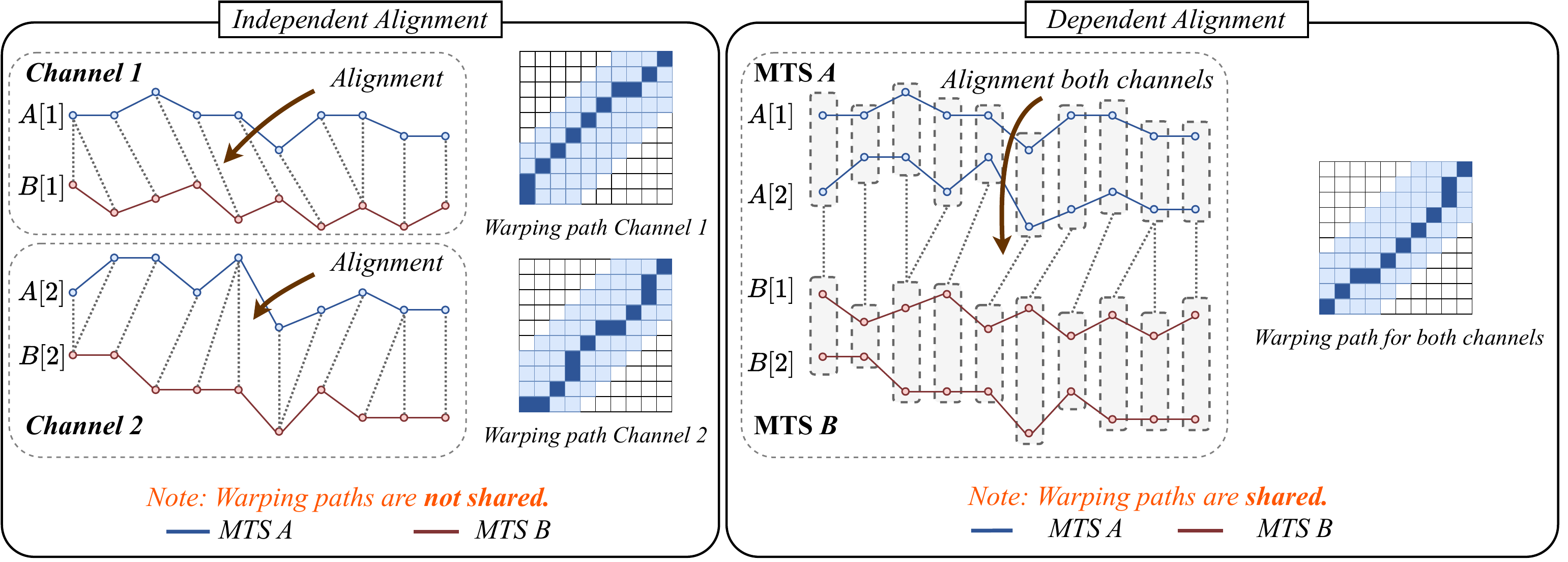}
    \caption{Visualization of the difference between the resulting warping paths between two MTS $\vec{A}$ and $\vec{B}$, when using dependent and independent to extend elastic measures to the multivariate case.}
    \label{fig:dep-indep}
 \vspace{-1.5em}    
\end{figure}

As discussed in Section~\ref{sec:multivariate_intro}, extending a univariate elastic measure in a \emph{channel-dependent} fashion involves finding a single alignment for the whole MTS, while \emph{channel-independent} extension involves finding an alignment for each pair of channels separately.
The difference between these dependency models for elastic measures is visualized in Figure~\ref{fig:dep-indep}.
For channel-dependent elastic measures, such an optimal global alignment is found by adopting the same general distance function of elastic measures discussed in Section~\ref{sec:elastic}, but replacing the internal distance function used to compare two time points (i.e., $D^v, D^h, D^d, D^u$) to a distance over two vectors with $c$ dimensions, rather than two scalars.
For channel-dependent DTW and its variants like DDTW, this implies that the squared Euclidean distance is used to compute the local distances, rather than the squared difference.
This results in the following expression for the local distance between two time points $\vec{X}_i$ and $\vec{Y}_j$ in a channel-dependent variant of DTW (referred to as DTW-D) as:
\begin{equation}
    \label{eqn:dtw_recursion}
    \footnotesize{
        D(i,j) = \left.
        \begin{cases}
            \sum_{k=1}^c (\vec{X}_i^{(k)} - \vec{Y}_j^{(k)})^2 & \text{if $i,j = 1$}\\
            D(i-1,j) + \sum_{k=1}^c (\vec{X}_i^{(k)} - \vec{Y}_j^{(k)})^2 & \text{if $i \neq 1$ and $j = 1$} \\
            D(i,j-1) + \sum_{k=1}^c (\vec{X}_i^{(k)} - \vec{Y}_j^{(k)})^2 & \text{if $i = 1$ and $j \neq 1$} \\
            min \begin{cases}D(i-1,j-1) + \sum_{k=1}^c (\vec{X}_i^{(k)} - \vec{Y}_j^{(k)})^2 \\ D(i-1,j) + \sum_{k=1}^c (\vec{X}_i^{(k)} - \vec{Y}_j^{(k)})^2 \\
            D(i,j-1) + \sum_{k=1}^c (\vec{X}_i^{(k)} - \vec{Y}_j^{(k)})^2 \end{cases} & \text{if $i, j \neq 1$}
        \end{cases}
        \right.
    }
\end{equation}
The same idea as DTW holds for other channel-dependent variants of edit-based measures like ERP, MSM, and TWED; the costs of each operation (i.e., insertion, deletion, substitution) are now based on vectors, rather than scalars. 
This means a summation is added over the channels at each local distance function $dist^D, dist^V, dist^H$.
The extension of MSM, however, comes with a caveat on the definitions of $dist^V(x_i,y_j)$ and $dist^H(x_i,y_j)$, that depend on the values of preceding points $x_{i-1}$ and $y_{j-1}$ being inside or outside the range $[x_i,y_j]$.
As such ranges are now defined by vectors, they become areas in $c$-dimensional space.
However, it is not straightforward to define what this 'critical area' should be. 
It could be a hypersphere surrounding both vectors, a hypercube around the vectors, or even an area between two hyperplanes orthogonal to a line that connects the two vectors, as done in SVMs.
In~\cite{shifaz23}, the authors argue for the first approach, and define the critical area as a hypersphere around the vectors, with the diameter being the Euclidean distance between the vectors.
While such intricacies do not exist for all elastic measures, this example does show that there is room for ambiguity in the definition of the channel-dependent extension of elastic measures.
Therefore, we leave exact definitions of the channel-dependent extension of all elastic measures to the user, and only present the general idea of the extension strategy here.

For LCSS, channel-dependent extension implies that multivariate subsequences are now compared, rather than univariate subsequences. Consequently, subsequences are deemed similar if they are similar in all channels.

Independent extension of elastic measures is straightforward; the distance between two MTS is computed as the sum of the distances between each pair of channels (e.g., the sum of the pairwise DTW distances).
To illustrate, the independent extension of DTW named DTW-I is defined as:
\begin{equation}
    DTW\text{-}I(\vec{X},\vec{Y}) = \sum_{k=1}^c DTW(\vec{X}^{(k)},\vec{Y}^{(k)})
\end{equation}


The choice between dependent and independent extension of elastic measures is tied to the assumed nature of time discrepancies that exist between two MTS, and how these discrepancies differ between the channels. 
For example, considering MTS sourcing from weather sensor arrays, the measurements between two MTS may be time-shifted due to a difference in the hardware or configuration of the arrays. In such a case, the offset is expected to be the same for all channels, and a dependent extension of the measure is appropriate.
In case the time discrepancies source from issues with individual sensors, such as latency or calibration issues, the time shifts are expected to differ between the channels, and an independent extension of the measure is appropriate.~\cite{yekta17,shifaz23}.

\vspace{-0.5em}
\subsection{Multivariate Sliding Measures}
The dependent extension of sliding measures involves simultaneously shifting all channels of one time series and calculating the distances between all possible shifts of that time series and another time series.
The channel-dependent extension of cross-correlations, named multivariate cross-correlations (MCC), can be computed with the use of 2-dimensional Fast Fourier Transforms (FFT2) as: 
\begin{equation}
    MCC(\vec{X},\vec{Y}) = IFFT2(FFT2(\vec{X}) * FFT2(\vec{Y})) 
\end{equation}
This function will output a \emph{matrix} containing the distance between $\vec{X}$ and all possible shifts of $\vec{Y}$ along both the time dimension and the dimension of channels.
Then to form a channel-dependent multivariate extension of SBD, the maximum \emph{normalized} cross-correlation across shifts on the time dimension is derived (considering only the shifts that cover \emph{all channels}\footnote{Note that this implies that we take the maximum of only one row of the matrix outputted by $MCC$.}).
Formally, the normalized cross-correlation for a shift $w$ is defined as: 
\begin{equation}
    MNCC_w(\vec{X},\vec{Y}) = \frac{MCC_w(\vec{X},\vec{Y})}{\sqrt{\sum_{j=1}^c \sum_{i=1}^n (\vec{X}_{i}^{(j)})^2} \cdot \sqrt{\sum_{j=1}^c \sum_{i=1}^n (\vec{Y}_{i}^{(j)})^2}}
\end{equation}
which sets up the definition of the channel-dependent variant of SBD as:
\begin{equation}
    SBD\text{-}D(\vec{X},\vec{Y}) = 1 - \max_w MNCC_w(\vec{X},\vec{Y})
\end{equation}
The dependent extension of other sliding measures can be defined similarly.

Independent extension of sliding measures is similar to that of elastic measures; the distance between two MTS is computed as the sum of the (sliding) distances between corresponding channels.
To illustrate, the independent extension of SBD named SBD-I is defined as:
\begin{equation}
    SBD\text{-}I(\vec{X},\vec{Y}) = \sum_{k=1}^c SBD(\vec{X}^{(k)},\vec{Y}^{(k)})
\end{equation}

The choice for the extension strategy is similar to that of elastic measures; it depends on the assumed nature of the time discrepancies over the channels.
For example, considering two MTS $\vec{X}$ and $\vec{Y}$ from sourcing from different weather sensor arrays, consider two scenarios: (a) the measurements in $\vec{Y}$ have a time delay of 1 minute compared to $\vec{X}$ due to the master clock of the array not being calibrated, and (b) the measurements of one sensor in $\vec{Y}$ are delayed by 1 second compared to the other sensors due to latency with the central processing unit of the array.
In the first case, the time discrepancy is the same for all channels, and a dependent extension of the measure is appropriate.
In the second case, the time discrepancy differs between the channels, and an independent extension of the measure is appropriate.

\vspace{-0.5em}
\subsection{Multivariate Kernel Measures}
As kernel measures are inherently functions over other distances, their extension strategy is inherited from the distance that is used as a base. For example, the dependent extension of SINK, named SINK-D, involves the usage of the dependent extension of NCC, while the independent extension of SINK, named SINK-I, involves the usage of the idea of independent extension.
As such, the SINK-D distance between two MTS $\vec{X}$ and $\vec{Y}$ is defined as:
\begin{align}
    k^D_s(\vec{X},\vec{Y}, \gamma) &= \sum_w e^{\gamma MNCC_w(\vec{X},\vec{Y})} \\ 
    SINK\text{-}D(\vec{X},\vec{Y},\gamma) &= \frac{k^D_s(\vec{X},\vec{Y}, \gamma)}{\sqrt{k^D_s(\vec{X},\vec{X}, \gamma) k^D_s(\vec{Y},\vec{Y}, \gamma)}}
\end{align}
Likewise, the SINK-I distance between two MTS $\vec{X}$ and $\vec{Y}$ is defined as:
\begin{align}
    SINK\text{-}I(\vec{X},\vec{Y},\gamma) = \sum_{k=1}^c SINK(\vec{X}^{(k)}, \vec{Y}^{(k)}, \gamma_k)
\end{align}
where $\gamma_k$ is the parameter value for channel $k$, and $SINK(\vec{X}^{(k)}, \vec{Y}^{(k)}, \gamma_k)$ denotes the univariate SINK measure for the channel.
This idea holds for all kernel measures, and as such, we will not provide explicit definitions for the MTS here.

\subsection{Multivariate Feature-based Measures}
Feature-based measures transform the original time series to a new representation in the form of a set of feature values, and then compute the distance between these feature representations using either a predefined function or a learned function, potentially in the form of a machine learning model~\cite{wang2006characteristic}.
Dependent extension of feature-based measures involves transforming the whole MTS into a new representation, while independent extension involves transforming each channel separately.
Particularly, channel-dependent extensions allow features (i.e., statistics) to be extracted with information from all channels, like the overall mean along with the mean per channel.
Channel-independent extensions, on the other hand, treat the channels as separate entities, meaning that only information about the channel itself can be used to extract features to represent the channel.
For example, when looking to use Characteristic-Based Clustering (CBC)~\cite{wang2006characteristic} on MTS, the dependent variant of CBC would involve computing global structural characteristics like trend and seasonality on the whole MTS, i.e., all channels should follow the same trend and seasonality for it to be considered in the distance computation.
The independent variant of CBC, on the other hand, would involve computing the trend and seasonality for each channel separately, meaning that a local trend or seasonality in one channel can also be considered in the comparison with other MTS.

\subsection{Multivariate Embedding Measures}
Dependent extension of embedding measures involves embedding the whole MTS into a new representation, while independent extension involves embedding each channel separately.
For example, the dependent extension of GRAIL involves creating an embedding of an MTS by computing the distance of that with other reference MTS, and then computing the Euclidean distance between the embeddings. The independent extension of GRAIL involves deriving an embedding for each channel by computing the distance of that channel with matching channels of other reference MTS. The final distance is then defined as the sum of the Euclidean distances between the embeddings of each channel.

Dependent extension of deep learning-based embedding measures involves training a deep neural network on the whole MTS, meaning that the model takes in all the channels as input, while still outputting a single embedding~\cite{yue2022ts2vec,meng2023mhccl,zerveas2021transformer}. 
This allows the model to learn the dependencies between the channels, which can be beneficial in cases where the processes that generate the channels are interdependent.
Independent extension of deep learning-based embedding measures involves training a deep neural network on each channel separately, meaning that the model takes in only one channel as input, and outputs a single embedding for that channel.
The final distance is then defined, for example,  as the sum of the Euclidean distances between the embeddings of each channel.
While this method does not allow the model to learn the dependencies between the channels, it can be beneficial in cases where the processes that generate the channels are independent.
If the models would take in the whole MTS as input in those cases, it could lead to model learning dependencies that do not exist, or even worse, to model learning dependencies that are not desired. 
Still, while there are arguments for an independent extension of a neural network-based measure, there currently exists no measure in the literature that can be categorized as such, to the best of our knowledge. There do, however, exist several dependent extensions such as Ts2Vec~\cite{yue2022ts2vec} and MHCCL~\cite{meng2023mhccl}.

In addition to the earlier discussed examples, there exist two specialized distances for MTS that can be categorized as multivariate embedding measures named the PCA similarity factor and Eros~\cite{yang04}. Both measures involve performing principal component analysis~\cite{jolliffe02} on each MTS, and then computing the sum of cosine similarities between the principal components.
The difference is that the PCA similarity factor sums the cosine similarities of all possible pairs of principal components, while Eros only considers the cosine similarities of matching ones (i.e., the diagonal of the similarity matrix) and takes a \emph{weighted} sum instead of a simple sum.
Particularly the PCA similarity factor is defined as:
\begin{equation}
    S_{PCA,k}(\vec{X},\vec{Y}) = \sum_{i=1}^k \sum_{j=1}^k cos^2(\omega_{i,j})
\end{equation}
where $\omega_{i,j}$ is the angle between the $i$-th and $j$-th principal component of $\vec{X}$ and $\vec{Y}$, and $k \leq v$ is the number of considered principal components. 
Eros, on the other hand, is defined as:
\begin{equation}
S_{Eros}(\vec{X},\vec{Y}) = \sum_{i=1}^c w_i | \cos(\omega_{i,i}) |
\end{equation}
where the weight vector $w$ is based on the (normalized) aggregated eigenvalues of a training set of MTS, as a measure of average importance.
Both measures transform the time series to a new representation, therefore they are categorized as embedding measures.
As PCA can only be performed on matrices (i.e., full MTS), both PCA Similarity Factor as Eros does not have a channel-independent variant of the measure; they are by definition channel-dependent.

\vspace{-1em}
\subsection{Multivariate Model-based Measures}
While certain (families of) models can only be used to describe univariate time series, many probabilistic models naturally extend to multivariate time series as well. Gaussian distributions, for example, can be defined on univariate data as well as on multivariate data. In the case of univariate data, the model is parameterized by a mean $\mu$ and a standard deviation $\sigma$. A multivariate model of $c$ variates, on the other hand, is defined by a vector of means $[\mu_1, \dots, \mu_c]$ as well as a covariance matrix $\Sigma = \bigl[\begin{smallmatrix} \sigma_{1,1} & \sigma_{1,2} & \dots  \\ \dots & \dots &  \\ \sigma{c,1} &  & \sigma{c,c} \end{smallmatrix}\bigr]$.
As HMMs are essentially collections of Gaussian distributions, they can also be trivially extended to the multivariate case by utilizing multivariate Gaussians.

As such, channel-dependent extensions of model-based measures involve building a model on the whole MTS, while channel-independent extensions involve building univariate models, and taking the sum of pairwise distances between these models. 
As some models are by definition univariate or multivariate, measures using these models may only have a channel-dependent or channel-independent variant.

\vspace{-1em}

%% file: section/S13_conclusion.tex
\section{Conclusion}
Distance measures play a crucial role in many time-series tasks, including clustering, classification, and so on. Additionally, each category in our taxonomy has distinct strengths and weaknesses. Comprehending these strengths and weaknesses and effectively applying suitable measures to specific tasks can be difficult. Fortunately, there exist some outstanding research papers that focus on evaluating distance measures, which provide important insights into these methods. \cite{ding2008querying} assess the effectiveness of 8 representation methods and 9 similarity measures, along with their variants, across 38 time-series datasets from various application domains, indicating that there does not exist a similarity measure that consistently outperforms the others. \cite{mirylenka} introduce 3 new correlation-aware measures and conduct an evaluation of distance measures using 43 real datasets. Their findings reveal that these proposed methods are capable of encoding information that the ED overlooks. Moreover, \cite{john2020debunking} explores the impact of 8 normalization methods and conduct a thorough evaluation of 71 time-series distance measures from 5 distinct categories using 128 time-series datasets \cite{dau2019ucr}. This study builds benchmark results for the time-series field and reveals 4 long-standing misunderstandings about time-series distance measures. This paper exhibits that the performance of measures can be improved by normalization methods other than z-score, there exist lock-step measures that can outperform ED, most elastic measures are unable to outperform sliding measures no matter what the supervision setting is, and the best elastic measure is not DTW.

In this survey, we comprehensively reviewed over 100 time-series distance measures and have classified them into 7 categories: lock-step measures, elastic measures, sliding measures, kernel measures, feature-based measures, model-based measures, and embedding measures. After defining each category in the taxonomy, we conduct a thorough discussion of the 7 defined categories for univariate time series, followed by a discussion on how univariate distance measures can be extended to the multivariate case.

To the best of our knowledge, this survey is the first to introduce a taxonomy with these 7 categories for distance measures. It offers the most extensive compilation of distance measures and delves into measures like sliding measures, which have not been thoroughly discussed in previous surveys. Furthermore, we conduct a detailed discussion and develop an extension strategy that broadens the concept of multivariate extension strategies for DTW proposed by Yekta et al.~\cite{yekta17}. This strategy is then applied to all 7 discussed categories. Through these contributions, we anticipate that our work will serve not only as a comprehensive reference but also as an inspiration for future research into distance measures and their associated applications.

As for possible future work, conducting a comprehensive and unbiased assessment of all the univariate time series collected in this survey is essential. This will offer a deeper understanding of the strengths and weaknesses of each measure and category. Moreover, while we have outlined a method for expanding univariate distance measures to multivariate cases for all 7 categories, there is still a need for thorough testing to evaluate the performance of these extended measures. Additionally, we hope our taxonomy serves as a guide and reference for classifying existing distance measures into proposed 7 categories. Finally, we are eager to witness the emergence of new distance measures that fit into our taxonomy, as well as additional measures that could further extend our taxonomy in the future.